\documentclass[showpacs,amsmath,
twocolumn,
aps,prb]{revtex4}
\usepackage{graphicx}
\usepackage{dcolumn}

\begin{document}

\title{Structural, electrical, and magneto-optical characterization of
  paramagnetic GaMnAs quantum wells} \author{M. Poggio} \author{R. C.
  Myers} \author{N. P. Stern} \author{A. C. Gossard} \author{D. D.
  Awschalom}

\affiliation{Center for Spintronics and Quantum Computation,
  University of California, Santa Barbara, CA 93106} 
\date{\today}

\begin{abstract}
  Growth of GaMnAs by molecular beam epitaxy is typically performed at
  low substrate temperatures ($\sim 250^{\circ}$C) and high As
  overpressures leading to the incorporation of excess As and Mn
  interstitials, which quench optical signals such as
  photoluminescence (PL).  We report on optical-quality Ga$_{1 -
    x}$Mn$_{x}$As/Al$_{0.4}$Ga$_{0.6}$As quantum wells (QWs) with $x <
  0.2$\% grown at a substrate temperature of $400^\circ$C.  Electrical
  and structural measurements demonstrate that this elevated
  temperature reduces As defects while allowing the substitutional
  incorporation of Mn into Ga sites. From a combination of Hall and
  secondary ion mass spectroscopy measurements we estimate that at
  least 70-90\% of the Mn incorporates substitutionally in all samples
  studied. The incorporation behavior shows both a substrate
  temperature and QW width dependence. The low defect density of these
  heterostructures, compared to typical lower temperature grown
  GaMnAs, enables the observation of both polarization-resolved PL and
  coherent electron spin dynamics, from which the conduction band
  exchange parameter is extracted. No evidence of long-range Mn spin
  coupling is observed, whereas negative effective Curie temperatures
  indicate spin heating due to photoexcitation. Light Mn-doping
  maximizes the electron spin lifetime indicating the importance of
  the Dyakonov-Perel decoherence mechanism in these structures. PL
  spectra reveal a low energy peak from shallow donors, which because
  of the paramagnetic behavior of its PL polarization, we ascribe to
  Mn interstitials.
\end{abstract} 

\maketitle

\section {Introduction}

Since the invention of the ferromagnetic semiconductor Ga$_{1 -
  x}$Mn$_{x}$As, by Ohno \textit{et al.},\cite{Ohno:1996} it has been
the subject of intense experimental work. The combination of
electronic and magnetic properties in this system has enabled many
interesting experimental demonstrations arising from the
carrier-mediated nature of its ferromagnetism,\cite{Dietl:2001} which
provides an excellent test bed for semiconductor spintronics since the
host material system is a highly engineered optoelectronic
semiconductor.\cite{Ohno:2004,Macdonald:2005} As a consequence of both
the low solubility of Mn in GaAs and the high concentrations of
substitutional Mn (Mn$_{Ga}$) required for ferromagnetism to occur
($\sim $5{\%}, $\sim $10$^{21 }$cm$^{ - 3})$, molecular beam epitaxy
(MBE) growth at present is performed at low substrate temperatures
($\sim 250^{\circ} $C ) and high arsenic overpressures. This is a
regime of growth in which defects, chiefly excess As and Mn
interstitials (Mn$_i$), are incorporated into the epilayers at
concentrations that quench sensitive optical properties, such as
photoluminescence (PL) and absorption. These optical properties can
provide direct measurements of the energy splitting for spin-up and
spin-down carriers at the band edges. In II-VI dilute magnetic
semiconductors (DMS), for example, polarization-resolved magneto-PL or
absorption have been used to extract the exchange
constants.\cite{Dietl:1994} The ability to perform such optical
measurements in GaMnAs is an important step in understanding the
exchange interactions in this material.

Polarization-resolved PL spectroscopy is a useful tool for the
measurement of conduction and valence band spin splittings in magnetic
fields. PL techniques including measurements of the exciton Zeeman
splitting, the Hanle effect, and the PL decay time have been used to
determine electron and hole g-factors in semiconductor
heterostructures.\cite{Snelling:1991,Snelling:1992,Traynor:1995}
Additionally, the measurement of Kerr rotation (KR) or Faraday
rotation (FR) serves as an extremely sensitive probe of magnetization,
resolving less than 10 electron spins in bulk GaAs.\cite{Kato3:2004}
This technique has led to a variety of advances in non-magnetic
GaAs-based heterostructures including the electrical tuning of the
g-factor,\cite{Salis:2001,Kato:2003,Poggio:2004} observation of strain
induced spin-orbit interaction,\cite{Kato1:2004,Kato2:2004} and the
measurement of the spin-Hall effect.\cite{Kato3:2004} In addition to
electron spin dynamics, the technique has been used to measure
magnetic ion spin coherence in II-VI ZnCdSe/MnSe quantum
structures,\cite{Crooker:1996} and recently, in experiments
demonstrating electrical control of the exchange enhanced electron
spin splitting.\cite{Myers1:2004} The growth of optical-quality GaMnAs
allows for the application of the aforementioned techniques and opens
the door to a variety of measurements including the precise
determination of the $s-d$ ($p-d$) exchange parameter, $N_0 \alpha$
($N_0 \beta$), between electron spins in the s-like conduction (p-like
valence) band and the 3d spins localized on Mn$^{2+}$. From a
practical standpoint, long electron spin lifetimes and compatibility
with previously developed heterostructures for electron spin control
favor the development of magnetically doped devices in GaAs-based
III-V materials.

The purpose of this work is to develop the capability to grow GaMnAs
structures by MBE in which coherent spin dynamics can be observed
optically. Preliminarily, our growth campaign focused on achieving
stoichiometric GaMnAs grown at a typical low temperature of
$250^{\circ}$C, at which the use of As overpressures leads to large
concentrations of excess As. Stoichiometric growth can be achieved at
low substrate temperatures by digital growth techniques, such as
atomic layer epitaxy (ALE) in which the As flux is shutter
controlled,\cite{Johnston:2003} or analog growth in which the As flux
is controlled by source temperature and/or valve position. The former
technique has enabled digital ferromagnetic heterostructures made up
of sub-monolayers of MnAs with independent control of charge carriers
in the non-magnetic GaAs spacer layers, while hybrid growths using
both high temperature MBE (optical layer) and low temperature ALE
(magnetic layer) have been developed to enable optical quality in
GaAs-AlGaAs quantum wells (QWs) with a ferromagnetic
barrier.\cite{Myers2:2004} Although polarization-resolved PL from
these QWs demonstrates a spin coupling between the magnetic layer and
carriers in the QWs, no time-resolved KR could be measured. In
previous work using the As-valve for flux control, it was found that
stoichiometric GaAs grown at $250^{\circ} $C could be achieved as
indicated by the low charge compensation of doped carriers (1$\times
$10$^{16}$ cm$^{ - 3})$; however, the incorporation of Mn at levels
required for ferromagnetism could only be achieved by using an As
overpressure. The resulting defects quenched both PL and time-resolved
KR signals.\cite{Unpublished:1} In the current work we investigate MBE
grown GaMnAs/AlGaAs QWs with low Mn-doping levels ($x < 0.2$\%). At
these Mn concentrations the substrate temperature can be increased to
$400^{\circ} $ C while allowing substitutional incorporation of Mn. In
contrast to low temperature stoichiometric growth, at the increased
growth temperatures used here an As overpressure does not result in
excess As incorporation, enabling the observation of PL and
time-resolved electron spin dynamics in GaMnAs QWs.

Here we discuss structural, electrical, and
magneto-optical properties with respect to the growth conditions and
incorporation behavior of Mn in GaMnAs QWs. The effect of the
substrate growth temperature (T$_{sub}$) on the incorporation of Mn is
studied by secondary ion mass spectroscopy (SIMS) measurements of the
Mn concentration profiles (Sec. III). The activation energy for Mn
acceptors in the QWs is found by measuring the hole concentrations as
a function of temperature using the Hall effect (Sec. IV). Comparison
between the SIMS and Hall data allow us to estimate the fraction of
Mn$_{Ga}$ acceptors incorporated in the QWs assuming that hole
compensation is dominated by Mn$_i$ donors (Sec. V). Time-resolved KR
measurements indicate that Mn doping $< 10^{19}$ cm$^{-3}$ maximizes
the electron spin lifetime. These data show evidence for spin heating
due to photoexcitation and no long-range Mn spin coupling. The
electron spin splitting increases with magnetic doping allowing for
the determination of the conduction band exchange constant $N_0\alpha$
(Sec. VI). This has led to the surprising observation in GaMnAs QWs of
an antiferromagnetic $N_0\alpha$, whose dependence on QW width
indicates a contribution from kinetic exchange due to the confinement
energy of the QW.\cite{Myers:2005} Lastly, we discuss the
polarization-resolved PL of the QWs in which the low energy peak is
attributed to the bound exciton emission from Mn$_i$ donors (Sec.
VII). The polarization of this PL is proportional to the magnetization
of Mn$_i$ within the QWs, providing an indirect optical readout of
magnetic moments within the QWs.

\section {Growth}

Samples are grown on GaAs substrates in a Varian GEN-II MBE system
manufactured by Veeco Instruments, Inc. In corroboration with the
recent findings of Wagenhuber \textit{et al.}\cite{Wagenhuber:2004},
we find that the inclusion of Mn growth capability in the MBE system
does not preclude the growth of high mobility samples. We have
measured low-temperature mobilities greater than $1.7\times 10^{6}$
cm$^{2}$/Vs in typical modulation Si-doped AlGaAs/GaAs two-dimensional
electron gas structures grown in our chamber at $630^{\circ} $C . For
Mn-doped QWs, however, we use a lower growth temperature to enable the
substitutional incorporation of Mn. Samples are grown at $400^{\circ}
$C as monitored and controlled during growth by absorption band edge
spectroscopy (ABES) using white-light transmission spectroscopy
through the substrate, providing a typical substrate temperature
stability of $\pm 2^{\circ} $C . The growth rate of GaAs is $\sim 0.7$
ML/sec and of Al$_{0.4}$Ga$_{0.6}$As is $\sim 1$ ML/sec as calibrated
by reflection high energy electron diffraction (RHEED) intensity
oscillations of the specular spot. The As$_{2}$:Ga beam flux ratio for
all samples is 19:1 as measured by the beam equivalent pressure of
each species using a bare ion gauge in the substrate position. Mn cell
temperatures for doping were extrapolated from growth rate
calibrations of MnAs measured at much higher growth rates using RHEED
oscillations; the actual, measured, value of Mn-doping concentrations
is discussed in detail in later sections.

\begin{figure}\includegraphics{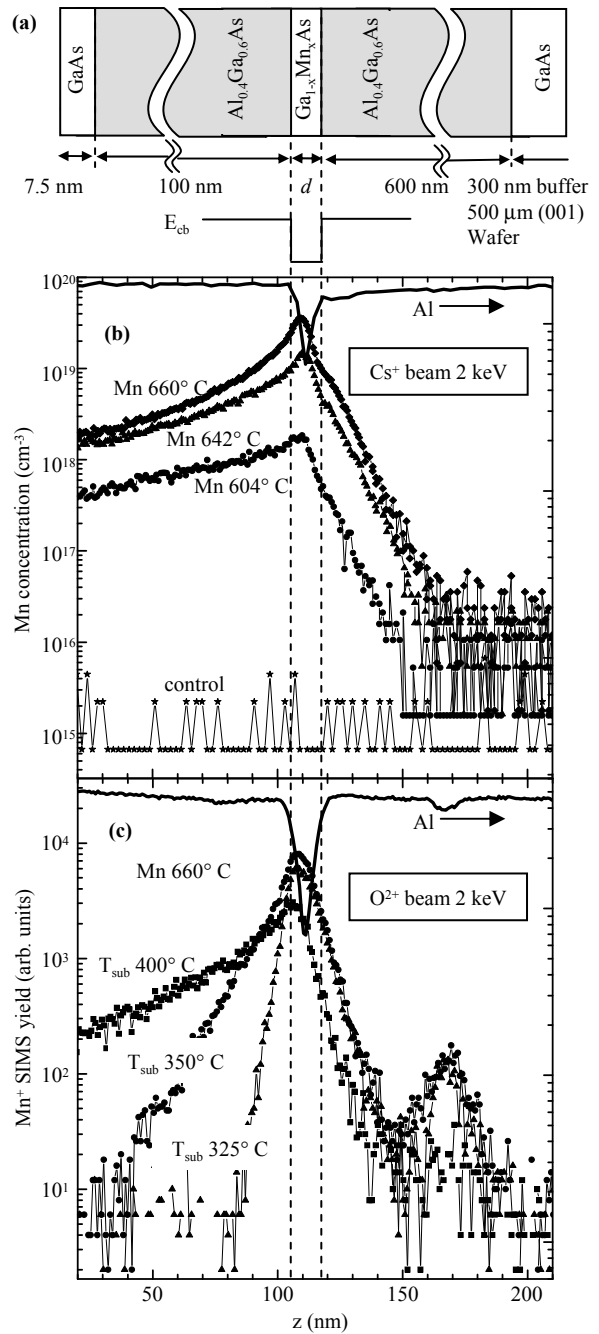}\caption{\label{fig1}
    (a) Schematic of sample layer structure and conduction band energy
    along the growth axis $\hat{z}$. (b) Mn concentration profile
    measured by SIMS for four 7.5-nm QWs with varying Mn-doping level
    (Mn effusion cell temperatures marked in the figure), and (c) for
    QWs with the same Mn doping level but varying substrate
    temperatures (marked in the figure). The un-calibrated Al SIMS
    signal (plotted as black lines) serve as markers for the QW
    region.  }\end{figure}

The QWs, shown schematically in Fig. \ref{fig1}(a), are grown on (001)
semi-insulating GaAs wafers using the following procedure.  The
substrate is heated to $635^{\circ} $C under an As overpressure for
oxide desorption and then cooled to $585^{\circ} $C . With the
substrate rotating at 10 RPM throughout the growth, a 300-nm GaAs
buffer layer is first grown using 5-s growth interrupts every 15 nm
for smoothing, which results in a streaky $2\times 4$ surface
reconstruction pattern as observed by RHEED. A 500-nm layer of
Al$_{0.4}$Ga$_{0.6}$As is grown followed by a 20-period digital
superlattice of 1-nm AlAs and 1.5-nm GaAs. The sample is then cooled
to the growth temperature (usually $400^{\circ} $C, but also
$350^{\circ} $C and $325^{\circ} $C) during which the RHEED pattern
changes to a $4\times 4$ reconstruction. The first QW barrier consists
of a 50-nm Al$_{0.4}$Ga$_{0.6}$As layer; during its growth the
$4\times 4$ pattern becomes faint and changes to $1\times 1$.  Before
the QW layer, a 10-s growth interrupt is performed to smooth the
interface; during this wait the RHEED partially recovers a $4\times 4$
reconstruction pattern. The GaMnAs QW layer deposition causes the
$4\times 4$ to again become faint leading to $1\times 1$, but during
the next 10-s wait on the top side of the QW, a $4\times 4$ partially
recovers. Increased Mn-doping leads to surface roughening as evidenced
by the development of a spotty RHEED pattern during and after the QW
growth. In contrast, lower-doped samples, which emit PL and show
time-resolved KR signal, display a streaky two-dimensional RHEED
pattern throughout their growth. The structure is completed by a top
QW barrier of 100-nm Al$_{0.4}$Ga$_{0.6}$As and a 7.5-nm GaAs cap
after which a streaky $4\times 4$ reconstruction pattern is observed.

\section {Secondary ion mass spectroscopy}

\subsection {Manganese-doping profiles}

The Mn concentration profile for each sample is measured using SIMS
and is plotted in Fig. \ref{fig1}(b) for a set of four 7.5-nm QWs
grown on the same day at $400^{\circ} $C with different Mn-doping
levels. The primary beam consists of Cs$^+$ ions with an impact energy
of 2 keV providing a depth resolution of 3.25 nm/e, as calibrated using
the atomically sharp AlGaAs/GaAs interface as a reference. The
secondary ion used to measure the Mn concentration is CsMn$^+$, which
is resilient to changes in ionization yield at the AlGaAs/GaAs
interface. The calibration of the SIMS Mn signal was performed using
Mn-ion implanted GaAs as a reference, while the change in ionization
yield of CsMn$^+$ between AlGaAs and GaAs was checked using Mn-ion
implanted AlGaAs reference samples. The z-axis calibration for the
SIMS scans was performed using the Al signal as a reference for the QW
region as well as the lower edge of the barrier, both layers grown at
low temperature. The depth of the crater could not be used to
calibrate the z-axis since it was found that the low temperature grown
layers sputter faster than the high temperature grown buffer layers.

The Mn concentration peaks near the center of the QW region, as
expected; but the surface side of the QW shows a large residual
concentration of Mn which incorporates into the structure even after
the Mn shutter has closed.  Such behavior is not unexpected, even if
the Mn-doping concentration is below the equilibrium solubility limit,
since the growing surface is not at equilibrium. This behavior has
previously been reported for similar structures in which Mn
$\delta$-doped GaAs grown at $400^{\circ} $C showed large surface
segregation.\cite{Nazmul:2003} We note that our results do not agree
quantitatively with those of Nazmul \textit{et al}, and we attribute
this discrepancy to the different methods used for substrate
temperature measurement in these two studies, noting that Mn
incorporation is highly sensitive to this growth parameter. In our MBE
system, the substrate temperature is measured directly by ABES, while
an indirect temperature reading from a radiatively coupled
thermocouple, used by Nazmul \textit{et al}, can be more than
$50^{\circ} $C from the actual substrate temperature.

Mn surface segregation during growth may lead to Mn clustering in our
samples and allow second phase magnetic precipitates, such as MnAs, to
form. We note, however, that though these impurities are likely present
in our samples, the schottky barrier around such precipitates prevents
their detection in the electrical or optical signal of free carriers
in the quantum wells. In the next section, the hole conductivity and
carrier densities measured in these samples indicate, in comparison
with the SIMS data, that most of the Mn impurities present in the
sample are substitutionally incorporated. Thus Mn surface segregation
and related growth defects have a negligible effect on our optical
studies of the exchange splittings.

\begin{figure*}\includegraphics[width=\textwidth]{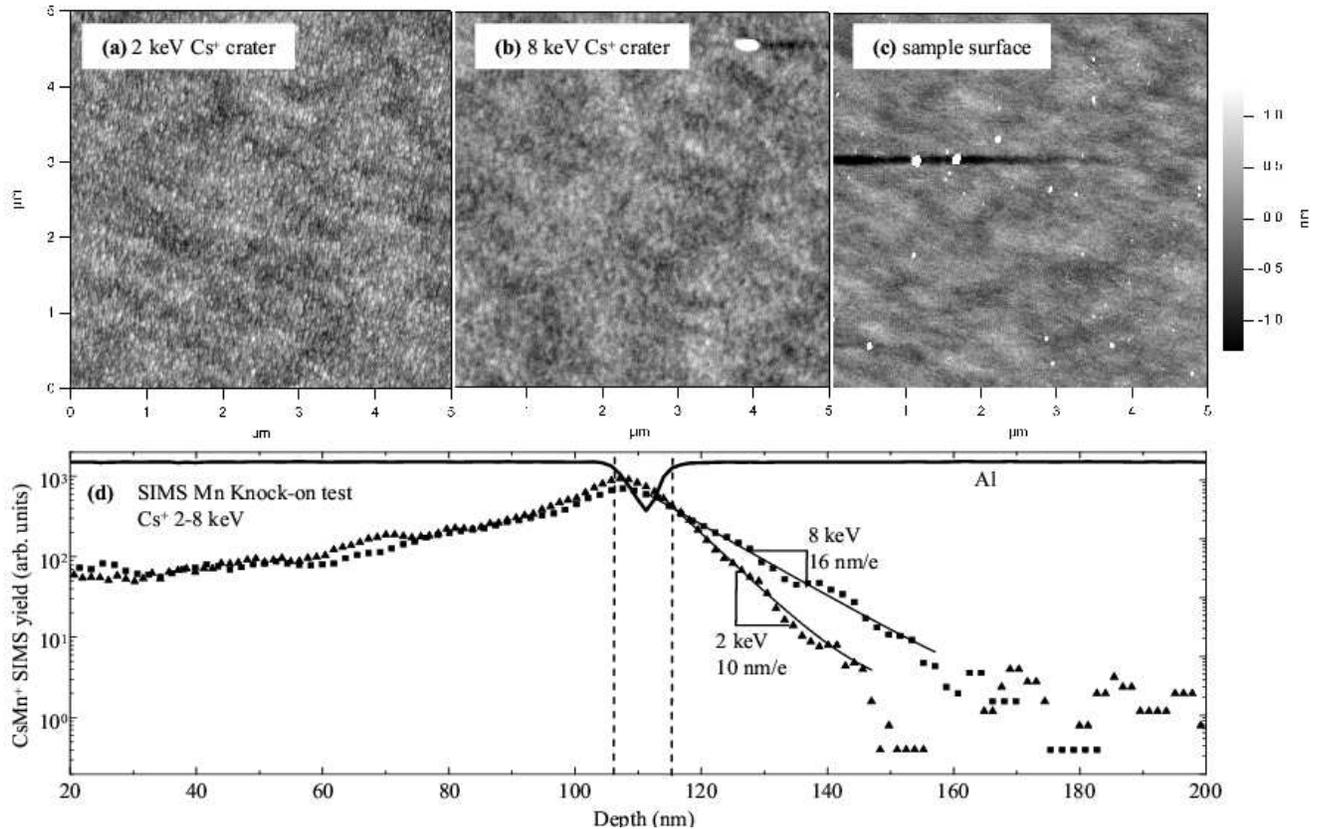}\caption{\label{fig19}
    Beam energy test for preferential Mn sputtering (knock-on). (a)
    and (b) show AFM images of the sample morphology in the center of
    SIMS craters formed by a 2 keV and 8 keV Cs$^+$ beam,
    respectively. (c) AFM image of the bare sample surface. (d) The Mn
    profiles in sample C measured at two different beam energies as
    labeled. Solid lines are exponential fits to the Mn profile tail
    below the QW.}\end{figure*}

Below the QW, the Mn concentration decreases toward the substrate
reaching a minimum point $\sim 100$ nm below the QW. The Mn profile is
broader than the Al depth profile indicating either a SIMS mesurement
artifact, such as poor depth resolution or preferential Mn sputtering,
i.e. knock-on effects, or that Mn diffuses into the barrier. The lack
of any temperature or doping level depedence on the Mn profile below
the QW eliminates the latter possibility, while Mn knock-on effect has
been reported in similar structure grown by Nazmul \textit{et al}. To
test for the presence of the knock-on effect, we run SIMS scans on the
same sample (C) at two different beam energies, 2 keV and 8 keV, Fig.
\ref{fig19}. Indeed, the Mn profile on the substrate side of the QW
shows a beam energy dependence with a decay of 17 nm/e for 8 keV and
10nm/e at 2 keV. The Mn tail on the surface-side of the QW, where Mn
incorporates as it floats along the surface, does not show any
significant dependence on the beam energy. As a further test, we use
an atomic force microscope (AFM) to measure the roughness of the SIMS
craters for the 2 keV and 8 keV scans as 0.32 nm RMS and 0.30 nm RMS,
respectively, while the roughness of the starting surface is 0.14 nm
RMS. Since the crater roughness does not depend on beam energy, we
conclude that the sample is uniformly sputtered at these two beam
energies. Therefore, the observed difference in decay of the Mn
profile is a result of the knock-on effect. Thus, the Mn profiles on
the substrate side of the QW are sharper than the SIMS data show.

After the lower Mn tail, $\sim 100$ nm below the QW, the Mn profile
becomes constant for all samples. Though the value of this background
Mn concentration is near the detection limit of SIMS, it is seen to
scale with Mn cell temperature as seen in Fig. \ref{fig1}(b). This
behavior suggests that Mn flux escapes from the hot Mn cell and
incorporates into the substrate even with the shutter closed. The Mn
background for the control sample, grown with a cold Mn cell, is well
below the SIMS detection limit.

\subsection {Substrate temperature dependence}

In Fig. \ref{fig1}(c), the Mn SIMS profiles are plotted for three
7.5-nm wide QWs grown with the same Mn cell temperature (same Mn beam
flux), but different substrate temperatures. These SIMS scans were
performed using an O$^{2+}$ 2 keV beam. Note that the Mn profiles
measured with the O$^{2+}$ beam show a large sensitivity to
AlGaAs/GaAs interfaces, probably due to oxygen impurities incorporated
during growth in the AlGaAs layers, thus the vertical axis of these
scans is uncalibrated (arbitrary units). The data, however, are
qualitatively meaningful in the interface free regions, e.g. the top
QW AlGaAs barrier.

As the substrate temperature is decreased, the Mn concentration
profiles become dramatically sharper, the surface tail is eliminated,
and the peaks become taller, indicating that, as expected, the Mn
incorporation is energetically more favorable at lower temperature.
Sharp Mn profiles, particularly on the surface side of the QW, are
desirable for the engineering of more complex magnetic quantum
structures in which a precise control of the magnetic doping is
required. Optical signals, however, even at the relatively high growth
temperatures of $325^\circ$C and $350^\circ$C, quench due to increased
defect densities even in the non-magnetic control samples. Preliminary
work shows that optical quality non-magnetic and magnetically doped
InGaAs/GaAs QWs can be grown at $350^\circ$C, suggesting that the loss
in signal for the GaAs/AlGaAs QWs is related to Al, well-known for its
impurity gettering of oxygen defects during MBE
growth.\cite{Akimoto:1986,Achtnich:1987} Also noteworthy is a
secondary peak in the Mn concentration 50 nm below the QW that occurs
only in the $325^\circ$C and $350^\circ$C grown samples. This Mn peak
corresponds to the interface between the high temperature and low
temperature grown QW barrier, a point at which a long growth pause
takes place. The peak indicates that the closed-shutter Mn flux, which
scales with Mn cell temperature as discussed previously, may
accumulate on the surface during the substrate cooling period before
low temperature growth begins, and may subsequently incorporate once
growth resumes, effectively delta doping the sample. In the
$400^\circ$C grown sample, grown on the same day and with the same Mn
cell temperature, this effective delta doping does not occur,
exemplifying the strong temperature dependence of Mn solubility in
GaAs. Another possible explanation for this peak is that background
impurities, such as oxygen, may incorporate during the long growth
pause leading to a change in the SIMS ionization yield. This
explanation is supported by the fact that the Al signal also shows a
change in intensity at this same region even though the Al
concentration should be constant.
 
\subsection {The effective Managanese concentration}

Since we probe carrier spin splittings at the band edges of the QWs,
then the presence of Mn in the AlGaAs barriers does not directly
interfere with our measurements. The leakage of Mn into the barriers,
however, makes the determination of the effective Mn concentration in
the QWs non-trivial. Since we are primarily interested in the
measurement of the conduction band exchange parameter, we define the
effective Mn concentration ($Mn$) in the QWs as the average of the
SIMS concentration profile $Mn(z)$ weighted by the electron
probability density along the growth axis $\rho_e (z)$. $\rho_e (z)$
is the square modulus of the ground state electron wave function
calculated using a one-dimensional Poisson-Schroedinger
solver.\cite{Snider:1} Thus, we calculate $Mn = \int{Mn (z) \rho_e (z)
  dz}$ by numerical integration. $Mn$ and the corresponding values of
$x$ are listed for a variety of QWs in Table \ref{table1}. These
values of $x$ are then used in the extraction of $N_0 \alpha$ from
time-resolved KR measurements.\cite{Myers:2005}

As discussed previously, the decay of the Mn profiles on the substrate
side of the QW is due to the knock-on effect, which is an artifact of
the SIMS measurement.Therefore, the calculation of $x$ as discussed
above contains this error. We estimate the uncertainty of $x$ by
calculating the effective concentration assuming that the Mn profile
on the substrate side of the QW is atomically sharp. Thus two values
of $x$ are calculated for each sample from which we calculate a
standard deviation in $x$. These errors are generally $<3$\%, except
for the $d=3$ nm QW set in which the error reaches 15\%. The $x$
errors lead to variations in our calculation of the $s-d$ exchange
parameter, which are included in the error bars of these parameters
(section VI).

\begin{table*}[htbp]
\caption{\label{table1} Quantum well substrate growth temperature ($T_{sub}$), width ($d$), room temperature hole density ($p$), Mn concentration ($Mn$), $x$, $p/Mn$, estimated fraction of substitutional Mn ($Mn_{Ga}/Mn$), and activation energy of holes (E$_a$).}

\begin{ruledtabular}
\begin{tabular}
{ccccccccc}
Sample& 
$T_{sub}$ ($^{\circ} $C) &
$d$ (nm) & 
$p$ (cm$^{ - 3}$)& 
$Mn$ (cm$^{ - 3}$) & 
$x$(\%)& 
$p/Mn$& 
$Mn_{Ga}/Mn$&
E$_a$ (meV)\\
\hline \\
A&
400& 
7.5& 
$1.02 \times 10^{18}$& 
$1.44 \times 10^{18}$& 
0.0065& 
0.71& 
0.90&
68\\
B&
400& 
7.5& 
$2.13 \times 10^{18}$& 
$1.19 \times 10^{19}$& 
0.0537& 
0.18& 
0.73&
73\\
C&
400& 
7.5& 
$4.10 \times 10^{18}$& 
$2.80 \times 10^{19}$& 
0.1266& 
0.15& 
0.72&
96\\
D&
400& 
3& 
-& 
$5.50 \times 10^{17}$& 
0.0025& 
-& 
0.67&
-\\
E&
400& 
3& 
$1.62 \times 10^{18}$& 
$2.60 \times 10^{18}$& 
0.0117& 
0.62& 
0.87&
-\\
F&
400& 
3& 
$3.37 \times 10^{18}$& 
$6.06 \times 10^{18}$& 
0.0274& 
0.56& 
0.85&
-\\
G&
400& 
10& 
$9.58 \times 10^{17}$& 
$2.97 \times 10^{18}$& 
0.0134& 
0.32& 
0.77&
54\\
H&
400& 
10& 
$2.15 \times 10^{18}$& 
$6.87 \times 10^{18}$& 
0.0310& 
0.31& 
0.77&
41\\
I&
400& 
10& 
$3.89 \times 10^{18}$& 
$1.78 \times 10^{19}$& 
0.0804& 
0.22& 
0.74&
46\\
J&
400& 
10& 
$4.61 \times 10^{18}$& 
$2.82 \times 10^{19}$& 
0.1274& 
0.16& 
0.72&
59\\
K&
400& 
5& 
$4.59 \times 10^{17}$& 
$9.65 \times 10^{17}$& 
0.0044& 
0.48& 
0.83&
-\\
L&
400& 
5& 
$1.47 \times 10^{18}$& 
$2.43 \times 10^{18}$& 
0.0110& 
0.61& 
0.87&
-\\
M&
400& 
5& 
$3.11 \times 10^{18}$& 
$7.03 \times 10^{18}$& 
0.0318& 
0.44& 
0.81&
36\\
N&
400& 
5& 
$4.01 \times 10^{18}$& 
$1.27 \times 10^{19}$& 
0.0574& 
0.32& 
0.77&
47\\
O&
350& 
7.5& 
-& 
-&
-& 
-& 
-&
-\\
P&
325& 
7.5& 
-& 
-& 
-& 
-& 
-&
-\\
Q&
400& 
500& 
$2.14 \times 10^{18}$& 
$1.35 \times 10^{19}$& 
0.0610& 
0.16& 
0.72&
71\\
R&
400& 
500& 
$6.86 \times 10^{18}$& 
$6.25 \times 10^{19}$& 
0.2823& 
0.11& 
0.70&
54\\
S&
400& 
500& 
$3.39 \times 10^{18}$& 
$1.39 \times 10^{20}$& 
0.6292& 
0.02& 
0.67&
53\\
\end{tabular}
\end{ruledtabular}
\label{tab1}
\end{table*}

\subsection {Manganese incorporation versus well thickness}
\begin{figure}\includegraphics{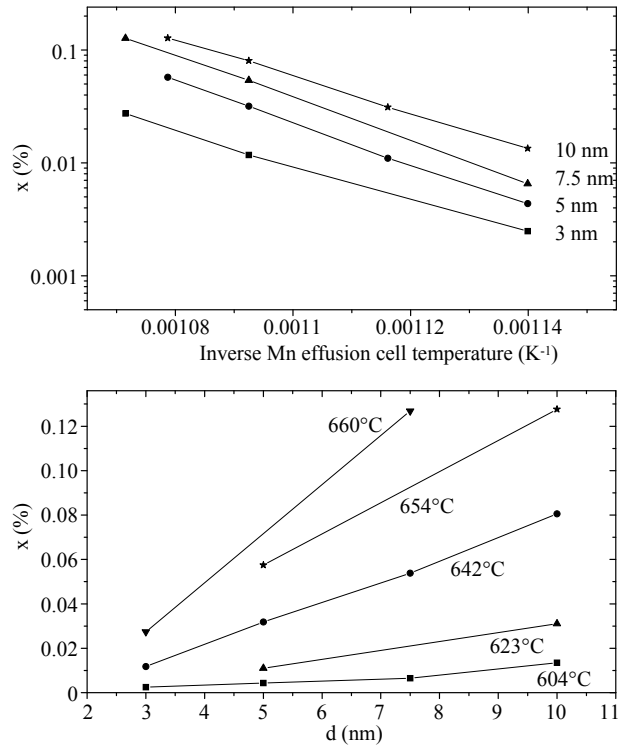}\caption{\label{fig2}
    (a) Percentage of $x$ as calculated from SIMS data (see text) as a
    function of the inverse Mn effusion cell temperature. Four quantum
    well sample sets of different well width are shown. (b) Percentage
    of $x$ versus quantum well width. Lines connect samples grown with
    the same Mn cell temperature, which is labeled in the
    figure.}\end{figure}

As discussed above, the Mn incorporation behavior in QWs is
non-trivial, showing a peak in its concentration profile above the
center of the well, and a strong dependence on substrate temperature.
Under these circumstances, we expect a strong variation of the
incorporation of Mn for wells of different widths under fixed Mn beam
flux. The Mn concentration ($x$) is plotted in Fig. \ref{fig2} (a) for
four QW sample sets of varying width as a function of the inverse Mn
effusion cell temperature. For each sample set of fixed width, the Mn
concentration increases exponentially with Mn cell temperature
indicating a linear relation between the Mn beam flux and doping
density. For any given Mn cell temperature, however, the doping
density decreases with QW width. This trend is explicitly plotted in
Fig. \ref{fig2} (b), where $x$ within each quantum well is plotted as
a function of quantum well width for various Mn cell temperatures.
This plot demonstrates that Mn incorporation is a strong function of
well thickness for the growth conditions used here, particularly for
high Mn beam fluxes (high effusion cell tempereatures).

The strong QW width dependence is likely due to the kinetics of Mn
incorporation. These results are not surprising considering that many
dopants do not imediately incorporate fully into semiconductors during
MBE growth, e.g. In is well-known to only fully incorporate into GaAs
once a critical density of In is deposited on the
surface,\cite{Toyoshima:1993} for this reason sharp InGaAs/GaAs QWs
are grown via predeposition of one monolayer of InAs followed by the
alloy growth and subsequent reevaporation of the ``In floating
layer''.\cite{Nagle:1993} In this case, it is expected that narrow
wells have lower total In densisties than wider wells, which is what
we observe in the case of the GaMnAs QWs.

\section {Hall effect}

Measurements of carrier concentration are carried out at $ T = 300$ K
on samples prepared in the Van der Pauw geometry and results are shown
in Table \ref{table1}. We examine carrier compensation related to As
defects by comparing the electron density of a sample of 1 $\mu $m of
n-GaAs (Si-doped $\sim 1 \times 10^{17}$ cm$^{ - 3}$) grown in typical
high temperature conditions ($580^{\circ} $C) to that of a sample with
the identical structure but grown at $400^{\circ}$C, the same
temperature as used for the Mn-doped QWs. In the $400^{\circ} $C grown
sample, the electron density is lower than in the $580^{\circ} $C
grown sample by $1.1 \times 10^{16}$ cm$^{ - 3}$, which provides an
estimate for the compensation due to non-Mn related growth defects,
i.e. excess As. Si incorporation is amphoteric which can lead to both
n and p doping (Si self-compensation), however, for high temperature
growth with Si doping level less than $5 \times 10^{18}$ cm$^{ - 3}$,
we expect only n-type doping from Si.\cite{Macguire:1987} At lower
temperature, Si may also act as an acceptor,\cite{Grandidier:1998}
thus compensation assumed to be due to As defects could actually be
occuring due to Si self-compensation leading to an overestimate of
excess As. For this reason, the carrier compensation we observe in the
$400^{\circ} $C grown sample serves as an upper limit on the amount of
excess As in our samples, $<1 \times 10^{16}$ cm$^{ - 3}$, which is
drastically smaller than the typical concentrations measured in low
temperature ($250^{\circ} $C) grown GaAs, $\sim 1 \times 10^{20}$
cm$^{ - 3}$.\cite{Missous:1994} We also note that the compensation
threshold for our samples is lower than for ALE grown n-GaAs, $2
\times 10^{18}$ cm$^{-3}$ .\cite{Johnston:2003}

\subsection {Manganese hole-doping calibration}

\begin{figure}\includegraphics{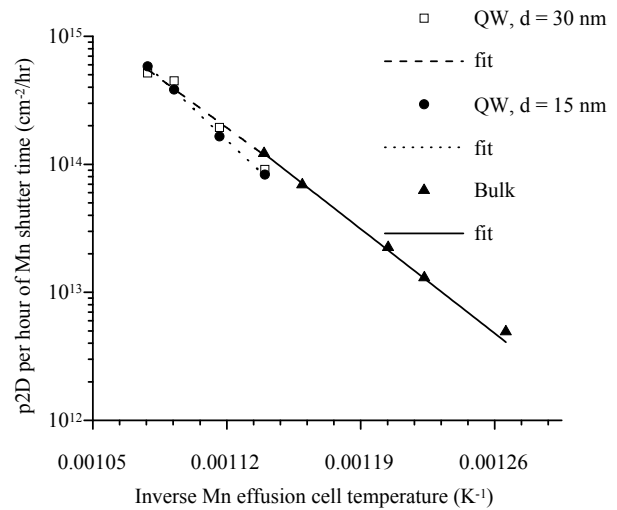}\caption{\label{fig3}
    The room temperature two-dimensional hole density (p2D) per hour
    (hr) of Mn shutter time as a function the inverse Mn source
    temperature. Three sample sets grown at $400^{\circ} $C are
    plotted and the data are fit to an exponential thermal activation
    (Arrhenius) equation, plotted as a line.  }\end{figure}

Given, as discussed above, that growth defects not related to
Mn-doping have a limited effect on the electronic properties for the
chosen growth conditions, then the hole density in Mn-doped samples
will be limited, for all practical purposes, only by the incorporation
behavior of Mn. In order to determine the nature of Mn incorporation
in our samples we investigate the p-doping dependence on Mn flux in a
bulk calibration sample series grown under the same conditions as the
Mn-doped QWs ($T_{sub} = 400^{\circ} $C). These samples consist of 1
$\mu $m of GaAs doped with Mn at various effusion cell temperatures.
We plot sheet concentration of holes (p2D) per hour of Mn shutter time
as a function of the inverse temperature of the Mn effusion cell in
these samples (Fig. \ref{fig3}). The doping rate demonstrates
exponential thermal activation, fitting well to an Arrhenius equation,
which indicates that there is a linear relation between the Mn beam
flux and the hole density. Thus, for this doping range and under these
growth conditions, Mn incorporation is mostly substitutional. The hole
densities for two additional sample sets, 15-nm and 30-nm GaMnAs QWs,
are plotted in Fig. \ref{fig3}. Arrhenius fits to these data match the
fit for the bulk sample set, which indicates that Mn$_{Ga}$
incorporation is sustained at these higher doping levels regardless of
heterostructure effects. This is surprising considering that the
effects of Mn penetration into the QW barriers and surface depletion
could modify electrical properties, particularly for the narrower
wells. The three-dimensional hole concentration ($p$) to $Mn$ ratio,
$p/Mn$, is listed in Table \ref{table1}. Fruitful comparison of these
values to those in typical GaMnAs ($x > 1$\% and $T_{sub} =
250^{\circ} $C) is difficult since GaMnAs usually contains orders of
magnitude higher Mn and As defect concentrations than the samples
discussed here.

\subsection {Manganese hole activation energy}

\begin{figure}\includegraphics{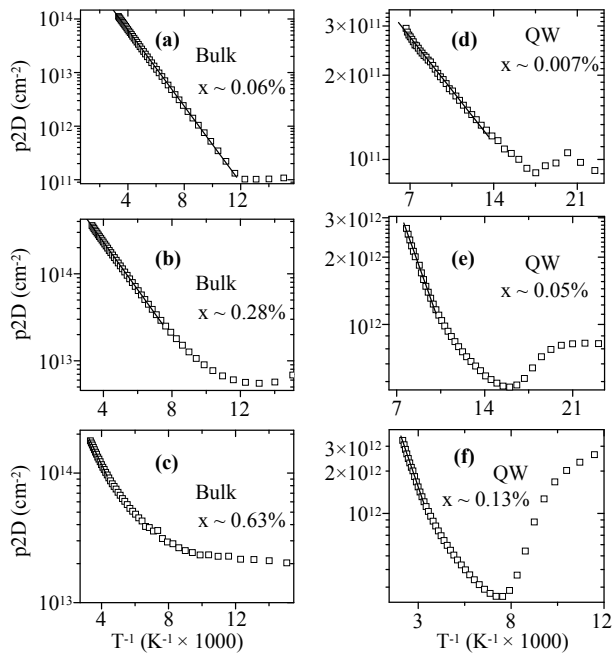}\caption{\label{fig4}
    The two-dimensional hole density (p2D) as a function of inverse
    sample temperature for 500-nm thick bulk GaMnAs (a-c) and for
    7.5-nm GaMnAs QWs (d-f). Exponential thermal activation
    (Arrhenius) fits are plotted as lines over the temperature range
    used for fitting.  }\end{figure}

In Fig. \ref{fig4}, the inverse sample temperature dependence of p2D
for a sample set of 500-nm thick layers of bulk GaMnAs and for a
sample set of 7.5-nm thick GaMnAs QWs are shown. At higher sample
temperatures the activation energy ($E_a$) for holes is extracted from
the linear portion of these plots, where the Mn-doped holes
demonstrate clear exponential (Arrhenius) thermal activation.  The
Arrhenius behavior is maintained over a larger temperature range in
the bulk than in the QWs, while at low temperature, carriers freeze
out in both samples. In both bulk and QW samples, the linear portion
of these plots decreases as the Mn concentration increases. Note that
the linear fits to the data are carried out over smaller temperature
ranges as $x$ increases. This behavior is expected to originate from
the formation and broadening of the Mn acceptor band for large doping
levels.

\begin{figure}\includegraphics{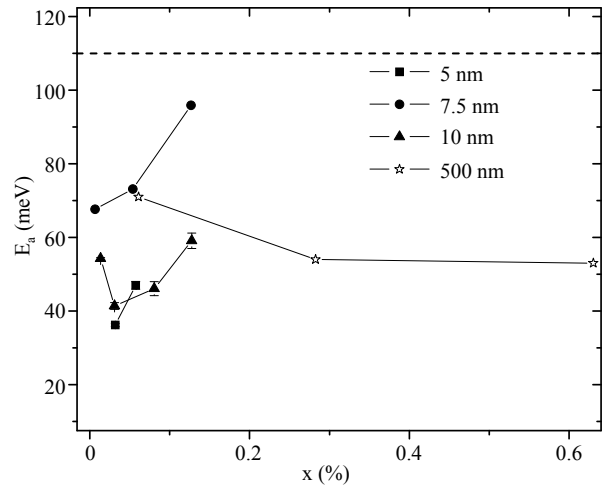}\caption{\label{fig5}
    The activation energy (E$_{a})$ for holes as a function of Mn
    concentration in bulk and in QW samples extracted from Arrhenius
    fits, e.g. Fig. \ref{fig4}. The 110-meV activation energy for an
    isolated Mn$^{2 + }$ ion in GaAs is plotted as a dashed line for
    reference.  }\end{figure}

In Fig. \ref{fig5}, the activation energies for a number of bulk and
QW samples are plotted as a function of Mn concentration. For all
samples measured, we have observed an activation energy lower than the
110 meV reported in the literature for an isolated Mn$_{Ga}$ in
GaAs.\cite{Yu:1979} This behavior was first observed in Mn doped GaAs
by Blakemore \textit{et al}, where the thermal activation energy of Mn
acceptors was found to be below the optical ionization energy of 110
meV.\cite{Blakemore:1973,Woodbury:1973} The activation energies were
lower than predicted due to impurity band broadening, suggesting a low
energy pathway due to sample inhomogeneity. In the case of impurity
band formation, we would expect a continued lowering of the activation
energy with increased Mn-doping as the Mn acceptor level broadens into
an impurity band, and eventually merges with the GaAs valence band
above the insulator to metal transition. For our bulk GaMnAs samples,
the activation energy does decrease with increased Mn doping, but for
QWs this effect is perhaps masked by a strong variation of the
temperature dependent hole concentration on QW width. For 3-nm wide
QWs, reliable activation energy for holes could not be extracted since
either p2D is not linear over any significant temperature range or the
hall data are too noisy to be reliable. This behavior is also observed
for the 5-nm wide QWs in which $E_a$ is only extracted for the two
highest doping levels, but is not reliable for the two lowest Mn doped
samples, whereas $E_a$ is reliably measured in all the 7.5-nm, 10-nm,
and bulk 500-nm samples. The dependence on well thickness is probably
due to carrier compensation from impurities, such as oxygen, that are
gettered by Al during the growth of the AlGaAs QW barriers, discussed
previously (see section IIIB). Barrier wave function penetration, and
therefore barrier defect compensation, increases as the well width
decreases.

\section {Substitutional Manganese Incorporation}

An estimate of the fraction of Mn which is substitutionally
incorporated can be made by assuming that each Mn$_{Ga}$ donates one
free hole and each Mn$_i$ compensates two holes.\cite{Erwin:2002} This
calculation has recently been used in ferromagnetic GaMnAs ($x \sim
1\%$) to estimate the concentrations of Mn$_i$ and Mn$_{Ga}$ with
respect to the crystal structure.\cite{Zhao:2005} Here we ignore As
defect compensation since it is negligible under our chosen growth
conditions ($<1 \times 10^{16}$ cm$^{ - 3}$). We also assume that
charge compensation due to Mn-defects other than Mn$_{i}$ (e.g. MnAs
precipitates) are negligible. Thus, we write,

\begin{equation}
\label{eq1}
p = Mn_{Ga}-2Mn_{i},
\end{equation}
 
\noindent
where $Mn_{Ga}$ is the concentration of substitutional Mn and $Mn_{i}$
is the concentration of interstitial Mn. Since $Mn$ is the sum of
$Mn_{Ga}$ and $Mn_{i}$, then we rewrite (\ref{eq1}) in terms of
$Mn_{Ga}/Mn$,

\begin{equation}
\label{eq2}
Mn_{Ga}/Mn = \frac{1}{3}(p/Mn)+\frac{2}{3}.
\end{equation}

\begin{figure}\includegraphics{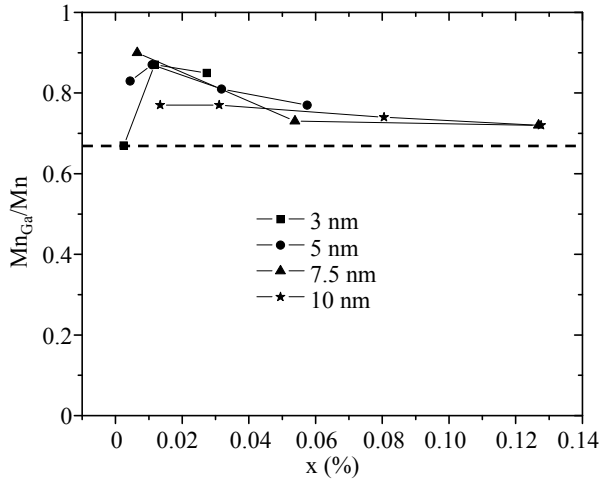}\caption{\label{fig6}
    The fraction of substitutional Mn ($Mn_{Ga}/Mn$) as a function of
    $x$, estimated assuming no surface depletion and full thermal
    activation of both Mn$_{Ga}$ doped holes and Mn$_{i}$ doped
    electrons. The minimum fraction of substitutional Mn (dashed line,
    2/3) for p-type GaMnAs is plotted.}\end{figure}

\noindent
Values of this ratio are provided for each sample in Table I and
plotted for four QW sample sets of varying doping density in Fig.
\ref{fig6}. The observation of hole conduction in all Mn doped QWs
provides a minimum value of $Mn_{Ga}/Mn = 2/3$, while the real value
is expected to be larger since Eq. (\ref{eq1}) ignores holes which
cannot be measured due to the incomplete thermal activation of
impurity bound holes at 300 K or surface depletion. For increased Mn-doping,
the $Mn_{Ga}/Mn$ ratio decreases indicating that incorporation of
growth defects, Mn$_i$ or Mn-containing second phases such as MnAs, is
becoming significant. Precipitates remove Mn from the lattice leading
to a reduction in hole concentration relative to the purely
substitutional case. Other forms of hole compensation such as
interstitial-substitutional pairs \cite{Yu:2002} and dimers of two
nearest neighbor substitutional Mn\cite{Raebiger:2004} are unlikely to
be present in our samples due to the low doping concentrations.
Optical signals, predominantly PL, show a similar degradation with
increased $Mn$ (see sections VI and VII). Note also, as indicated by
the temperature dependence of the hole concentration, that
particularly in the narrow QWs, $d=3$ and 5 nm compensation of holes
due to barrier defects may also deflate our estimate of $Mn_{Ga}/Mn$.
Despite these factors the fraction of Mn$_{Ga}$ in all our samples is
approximately 70-90\% as shown in Fig. \ref{fig6}.

\section {Time-Resolved Kerr Rotation}

Electron spin dynamics are measured by time-resolved KR with the
optical axis perpendicular to the applied magnetic field $B$ (Voigt
geometry) and parallel to the growth direction $\hat{z}$. The
measurement, which monitors small rotations in the linear polarization
of laser light reflected off of the sample, is sensitive to the spin
polarization of electrons in the conduction band of the
QW.\cite{Crooker:1996} A mode-locked Ti:Sapphire laser with a 76-MHz
repetition rate and 250-fs pulse width tuned to a laser energy $E_L$
near the QW absorption energy is split into a pump (probe) beam with
an average power of 2 mW (0.1 mW).  The helicity of the pump beam
polarization is modulated at 40 kHz by a photo-elastic modulator,
while the intensity of the linearly polarized probe beam is modulated
by an optical chopper at 1 kHz for lock-in detection.  Both beams are
focused to an overlapping 50-$\mu$m diameter spot on the sample which
is mounted within a magneto-optical cryostat. The time delay $\Delta
t$ between pump and probe pulses is controlled using a mechanical
delay line. The pump injects electron spins polarized perpendicular to
$B$ into the conduction band of the QW. The change in the probe
polarization angle, $\theta _K \left( {\Delta t} \right)$ is
proportional to the average electron spin polarization in the QW and
is well fit to a single decaying cosine, $\theta _K \left( {\Delta t}
\right) = \theta_{\perp} e^{ - \Delta t / T_2 ^\ast }\cos \left( {2\pi
    \nu _L \Delta t + \phi } \right)$, where $\theta_{\perp}$ is
proportional to the total spin injected, $T_2 ^\ast$ is the
inhomogeneous transverse spin lifetime, $\nu _L$ is the electron spin
precession (Larmor) frequency, and $\phi$ is the phase offset. No
evidence of Mn$^{2+}$ spin precession, which occurs in II-VI
magnetically doped QWs,\cite{Crooker:1996} has been observed in the
samples studied here.  The order of magnitude smaller $x$ in our III-V
QWs compared to the II-VI QWs puts any Mn$^{2+}$ spin precession
signal below the experimental detection limit.

\begin{figure}\includegraphics{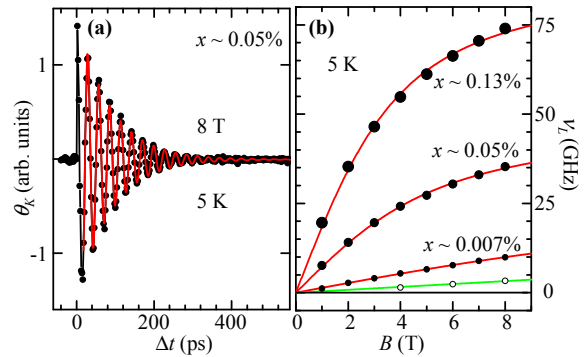}\caption{\label{fig7}
    (Color) Time-resolved electron spin dynamics in $d=7.5$ nm GaMnAs
    QWs. (a) An example of KR data (points) together with fit (line).
    (b) $\nu_L$ as a function of $B$ for different $x$ values (solid
    points); larger points indicate increasing $x$. Open data points
    are for the $x=0$ sample. Red lines in (b) are fits to Eq.
    (\ref{eq5}).  }\end{figure}

Fig. \ref{fig7}(a) shows typical time-resolved KR data measured at $B
= 8$ T for a Mn-doped QW ($d=7.5$ nm and $x\sim0.05\%$) together with
fit, as described above, demonstrating electron spin coherence in the
GaMnAs system. KR data showing electron spin precession was observed
to quench in all samples for $x > 0.13$\%.

\subsection{Transverse electron spin lifetime}

\begin{figure}\includegraphics{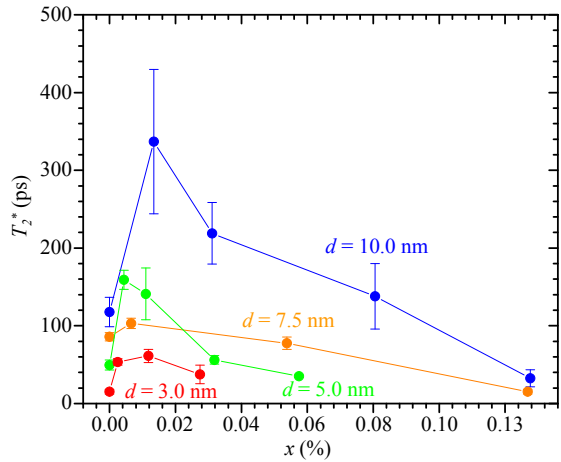}\caption{\label{fig8}
    (Color) The transverse electron spin lifetime ($T_2 ^\ast$) at $T
    = 5$ K versus the percentage of Mn for four quantum well sample
    sets of varies width. The plotted values of $T_2 ^\ast$ are the
    mean values from 0 to 8 T and the error bars represent the
    standard deviations.}\end{figure}

The transverse electron spin lifetime ($T_2 ^\ast$) is plotted in Fig.
\ref{fig8} as a function of the percentage of Mn for all four QW
sample sets. In all samples measured, we observe an increase in $T_2
^\ast$ with Mn doping as compared to the un-doped control samples.
This increase is consistent with the D'Yakonov-Perel (DP) spin
relaxation mechanism since increasing impurity concentration makes the
process of motional narrowing more efficient by providing additional
momentum scatters.\cite{Fabian:1999} After reaching a maximum at very
low Mn-doping ($x \sim 0.01$\%), $T_2 ^\ast$ drops off as a function
of $x$ as shown in Fig. \ref{fig8}. This behavior suggests that for $x
> 0.01$\%, the DP mechanism is no longer dominant. In this regime
either the Elliot-Yafet (EY) or the Bir-Aronov-Pikus (BAP) relaxation
mechanisms may limit conduction electron spin lifetimes, since both
should increase in strength with increasing $x$.\cite{Optical:1984} EY
relaxation, due to the spin-orbit interaction, grows stronger with
larger impurity concentration while, the BAP process, based on the
electron-hole exchange interaction, increases with increasing hole
doping.

In this discussion we have so far ignored the effects of the $s-d$
exchange interaction on the electron spin relaxation process. In II-VI
DMS, the presence of magnetic impurities leads to large relaxation
rates limiting the conduction electron spin lifetime.
\cite{Crooker:1996} Magnetic impurity doping in these materials
results in relaxation through spin-flip scattering arising from the
$s-d$ exchange interaction. While the samples discussed in this report
have $s-d$ exchange energies which are several orders of magnitude
smaller than in typical II-VI DMS, the effect of magnetic interactions
on $T_2^\ast$ cannot be totally ruled out. Several experiments can be
carried out in order to examine the role of exchange interactions in
the decoherence of electron spin including a finer dependence of
$T_2^\ast$ on $x$ and a dependence on the temperature. A detailed
study of changes in $T_2^\ast$ as a function of QW width also may
discern between the DP and the exchange scattering
mechanism.\cite{Semenov:2003} Such detailed studies will be the
subject of future work. Finally we note that qualitatively similar
dependence of $T_2^\ast$ on Mn-doping as shown in Fig. \ref{fig8} has
recently been observed in InGaMnAs/GaAs QWs.\cite{Unpublished:1}

\subsection{Larmor precession frequency}

$\nu _L$ is proportional to the total conduction band spin splitting
between spin-up and spin-down electrons ($\Delta
E=E\uparrow-E\downarrow$) and can be expressed in terms of the Zeeman
splitting ($\Delta E _g$), and the $s-d$ exchange splitting ($\Delta
E_{s - d}$):
\begin{equation}
\label{eq4}
h \nu_L = \Delta E = \Delta E _g + \Delta E_{s-d} = 
g _e \mu _B B - x N _0 \alpha \langle S_x \rangle.
\end{equation}
Here $h$ is Planck's constant, $g _e$ is the in-plane electron
g-factor, $\mu _ B$ is the Bohr magneton, and $\langle S_x \rangle$ is
the component of Mn$^{2+}$ spin along $B$.  $\langle S_x \rangle = -
\frac{5}{2} B_{5/2} \left ( \frac{5 g_{Mn} \mu_B B}{2 k_B
    (T-\theta_P)} \right )$, where $B_{5/2}$ is the spin-5/2 Brillouin
function, $g_{Mn}$ is the g-factor for Mn$^{2+}$, $k_{B}$ is
Boltzman's constant, and $\theta_P$ is the paramagnetic Curie
temperature. Note that since the g-factor for Mn$^{2+}$ ($g_{Mn}=2$)
is positive, for $B>0$, then $\langle S_x \rangle < 0$. We emphasize
that a measurement of $\nu_L$ alone, because of phase ambiguity, does
not determine the sign of $\Delta E$.

\begin{figure}\includegraphics{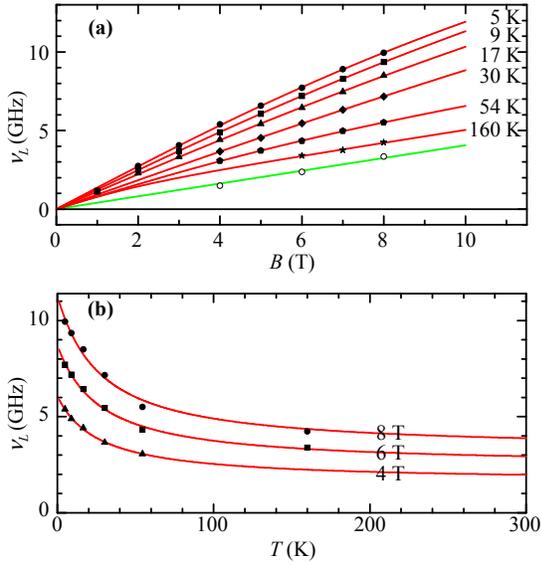}\caption{\label{fig9}
    (Color) Temperature dependence of $\nu_L$ for 7.5-nm quantum
    wells. (a) The effect of increasing $T$ on the $B$ dependence of
    $\nu_L$ for the sample with $x\sim0.007\%$ (solid points) and for
    the $x=0$ sample (open points). (b) $T$ dependence of $\nu_L$ at
    constant $B$ for the $x\sim0.007\%$ sample. Red lines in (a) and
    (b) are fits to Eq.  (\ref{eq5}).  }\end{figure}

In Fig. \ref{fig7}(b), $\nu_L$ is plotted as a function of $B$ for a
set of four samples with $d=7.5$ nm and varying $x$.  The non-magnetic
($x=0$) sample shows a linear field dependence of $\nu_L$, from which
we extract values of $g_e$ as described in Eq. (\ref{eq4}). As the Mn
doping concentration is increased, $\nu_L$ increases and its $B$
dependence becomes non-linear. Further, this field dependence shows
the same Brillouin function behavior that is expected for the
magnetization of paramagnetic GaMnAs, Eq. (\ref{eq4}). The dependence
of $\nu_L$ on $B$ and $T$ for the sample with $d = 7.5$ nm and
$x\sim0.007\%$ is plotted in Fig.  \ref{fig9}(a) and (b) together with
values for the control sample, $x=0$ and $d=7.5$ nm. For the magnetic
sample, as $T$ is increased, $\nu_L$ decreases asymptotically toward
the control sample value $g_e \mu_B B / h$ without crossing zero (Fig.
\ref{fig9}(a)). Thus, it follows from Eq.  (\ref{eq4}) and from the
sign of $\langle S_x \rangle$ that for $d = 7.5$ nm, $N_0 \alpha$ has
the same sign as $g_e$.  For $d = 7.5$ nm, $g_e <
0$,\cite{Snelling:1991} and thus $N_0 \alpha < 0$. This conclusion is
also supported by the QW width dependence discussed below.

Using $g_e$ extracted from the $x = 0$ sample (green line) and Eq.
(\ref{eq4}) we fit $\nu_L$ data as a function of $B$ and $T$ to,
\begin{equation}
\label{eq5}
\nu_L = \frac{g _e \mu _B B}{h} +\frac{5 A}{2 h} B_{5/2} \left ( \frac{5 \mu_B B}{k_B (T - \theta_P)} \right ),
\end{equation}
which has only two fit parameters, $A$ and $\theta_P$. Comparing Eqs.
(\ref{eq5}) to (\ref{eq4}), it is clear that $ A = x N_0 \alpha $. The
data in Fig. \ref{fig7}(b) and Fig. \ref{fig9} are fit to Eq.
(\ref{eq5}), with fits shown as red lines.  A large negative
$\theta_P$ (-24 K) is extracted from the fits for the sample with the
lowest Mn doping (Fig. \ref{fig9}), which may be explained by an
increased spin temperature of Mn$^{2+}$ due to photoexcitation. This
effect has been reported in II-VI DMS for low magnetic doping
levels.\cite{Keller:2001} Also supporting this hypothesis, we find
smaller values of $|\theta_P|$ ($<7$ K) in samples with larger $x$.

\begin{figure}\includegraphics{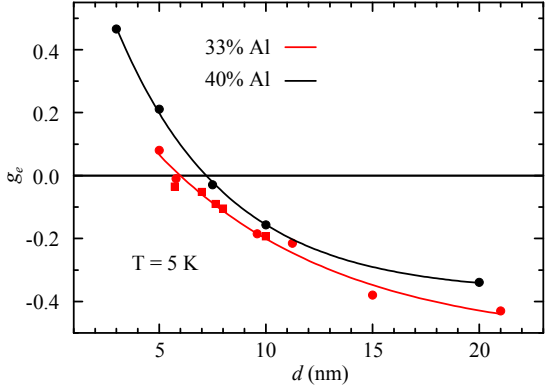}\caption{\label{fig10}
    (Color) $g_e$ as a function of $d$. Black points are from control
    ($x=0$) samples of this study, red circles and squares are from
    references,\cite{Snelling:1991,Poggio:2004} respectively. Lines
    guide the eye. 
}\end{figure} 

\begin{figure}\includegraphics{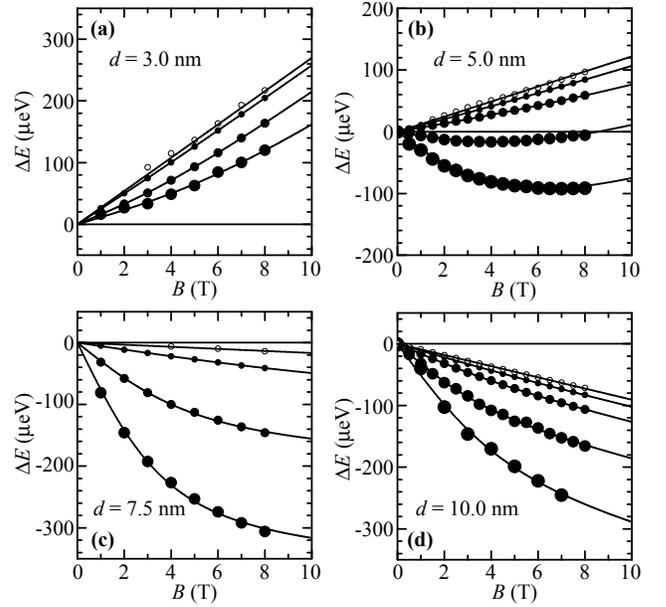}\caption{\label{fig11}
    $\Delta E$ as a function of $B$ at $T = 5$ K for QWs with (solid
    circles) and without Mn doping (open circles); larger points
    indicate increasing $x$. Fits to Eq. (\ref{eq5}) are shown as
    lines.  }\end{figure}

$N_0 \alpha$ is examined in detail for QWs of varying $d$.  For this
analysis, we examine four sets of samples with various $x$ (including
$x = 0$) for $d = 3, 5, 7.5$ and $10$ nm. Note that each sample set of
constant $d$ was grown on the same day, which we have observed to
reduce QW thickness variations between samples within each set from
$\sim3$\% to $<1$\%. Variations in QW thickness can affect the
electron g-factor and therefore result in errors in the determination
of $x N_0 \alpha$. By growing samples on the same day, the error in
the determination of $x N_0 \alpha$ is reduced from 10\% to less than
3\% as compared with samples grown on different days. In Fig.
\ref{fig10}, $g_e$ in the non-magnetic ($x=0$) QWs is plotted as a
function of $d$ together with data from two other
publications.\cite{Snelling:1991,Poggio:2004} Our data track the
thickness dependence of the QW g-factor as previously reported with a
slight positive shift in $g_e$.  The larger Al concentration (40\%) in
the QW barriers used in our samples versus the concentration (33\%)
used in Refs. \cite{Snelling:1991,Poggio:2004} accounts for this
discrepancy.\cite{Weisbuch:1977} Knowing the absolute sign of $g_e$
for QWs of any width, we determine the sign of $N_0 \alpha$ for each
$d$ in the manner described previously. With a calibrated sign,
$\Delta E = h \nu_L$ is plotted in Fig.  \ref{fig11} as a function of
$B$ for all four QW sample sets with varying $d$.  As shown in Fig.
\ref{fig11}, for any given $d$, $\Delta E$ decreases as $x$ increases.
 Following from Eq. (\ref{eq4}) and from the
sign of $\langle S_x \rangle$, this demonstrates that $N_0 \alpha$ is
negative, i.e.  antiferromagnetic, which has been reproduced
unambiguously in over 20 additional samples. A negative $N_0 \alpha$
has also been measured in recent time-resolved FR measurements in
InGaMnAs/GaAs QWs.\cite{Unpublished:1}

\begin{figure}\includegraphics{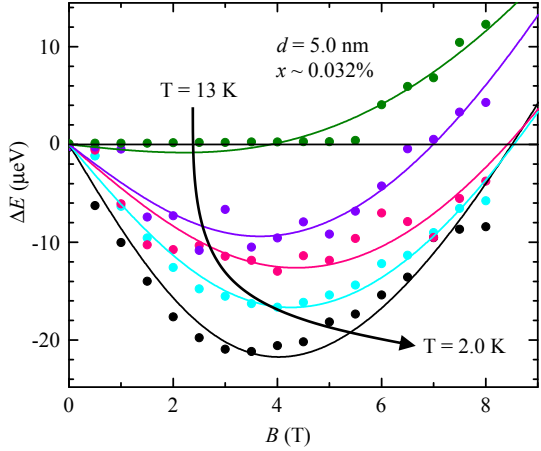}\caption{\label{fig12}
    (Color) $\Delta E$ for the sample with $d=5$ nm and $x\sim0.032\%$
    at various temperatures. Fits to Eq. (\ref{eq5}) are shown as
    lines.  }\end{figure}

The effect of increasing temperature on the $B$ dependence of $\Delta
E$ for the $d=5$ nm and $x\sim0.032\%$ sample is shown in Fig.
\ref{fig12}, which dramatically illustrates the negative $s-d$
constant. For $ d = 5$ nm, $g_e$ is weakly positive, thus for $B>0$
and at high temperature $\Delta E > 0$. As the temperature descreases,
$\Delta E_{s-d}$ becomes more negative as the paramagnetic
susceptibility increases. At $T = 10$ K and $B = 7$ T, $\Delta E = 0$
since the $s-d$ exchange splitting is equal and opposite to the Zeeman
splitting. For lower temperature, $\Delta E < 0$ since $|\Delta
E_{s-d}| > |\Delta E _g|$. We note that the data are well fit to Eq.
(\ref{eq5}) despite their highly non-linear nature. We contrast our
observation of antiferromagnetic $s-d$ exchange in III-V GaMnAs, with
the ferromagnetic $s-d$ exchange ubiquitous in II-VI DMS. In those
systems symmetry forbids hybridization of $s$ and $d$ orbitals, such
that only direct (ferromagnetic) $s-d$ exchange is
possible.\cite{Larson:1988} The antiferromagnetic $s-d$ exchange in
GaMnAs may be due to the narrower band gap of this material compared
with II-VI, such that the conduction band has partial p-character thus
allowing hybridization with the d orbitals localized on the Mn$^{2+}$
impurities.

\begin{figure}\includegraphics{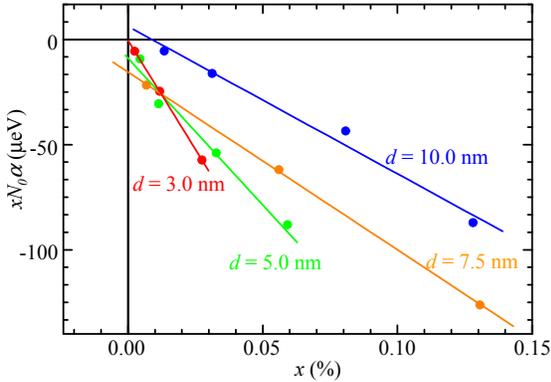}\caption{\label{fig13}
    (Color) $x N_0 \alpha$ as a function of $x$ from fits shown in
    Fig.  \ref{fig11}; error bars are the size of the points. Linear
    fits are shown for each sample set of constant $d$.  }\end{figure}

\begin{figure}\includegraphics{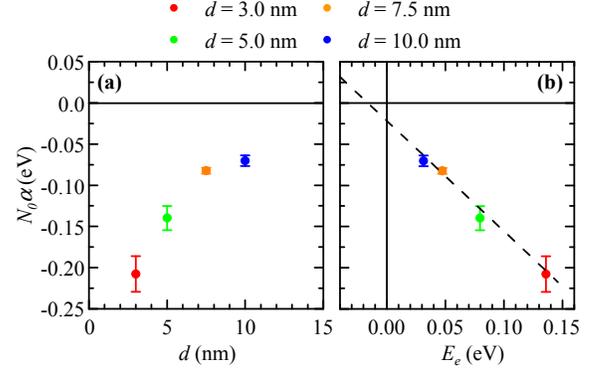}\caption{\label{fig14}
    (Color) The effect of confinement on $s-d$ exchange coupling. (a)
    $N_0 \alpha$ extracted from fits in Fig. \ref{fig13} and plotted
    as a function of $d$. (b) $N_0 \alpha$ as a function of electron
    kinetic energy for GaMnAs.  }\end{figure}

In Fig. \ref{fig13}, the fit parameter $A = x N_0 \alpha$ is plotted
as a function of $x$ together with linear fits for each sample set of
constant $d$. The finite values of $x N_0 \alpha$ at $x = 0$,
extrapolated from the linear fits, are attributed to either the
experimental error in the determination of $g_e$ in the non-magnetic
QWs or error in the measurement of $x$, both of which have a
negligible effect on the slope.  These linear fits demonstrate that
$N_0 \alpha$ is constant over the measured doping range for QWs with
the same width, but it varies with $d$ as plotted in Fig.
\ref{fig14}(a). $N_0 \alpha$ is more negative the narrower the QW,
while it appears to saturate for wide QWs.  In II-VI DMS QWs, a
negative change in $N_0 \alpha$ as large as $-170$ meV was previously
reported for increasing confinement and was attributed to a kinetic
exchange coupling due to the admixture of valence and conduction band
wave functions.\cite{Merkulov:1999}

We plot $N_0 \alpha$ as a function of the electron kinetic energy
($E_{e}$) in Fig. \ref{fig14}(b), and the data are linear. Here, $E_e$
is defined as the energy between the bottom of the conduction band in
the GaAs QW and the ground state energy, which is calculated using a
one-dimensional Poisson-Schroedinger solver and the material and
structural parameters of the QWs.\cite{Snider:1} Extrapolating to
$E_{e}=0$ we obtain a bulk value of $N_0 \alpha=-23\pm8$ meV for
GaMnAs. A change in $N_0 \alpha$ as large as $-185$ meV is observed in
the narrowest wells measured ($d=3$ nm) and the slope of $N_0
\alpha(E_{e})$ is roughly the same as reported by Merkulov \textit{et
  al.} in II-VI DMS. Since $N_0 \alpha > 0$ in bulk II-VI DMS, the
kinetic exchange effect appears as a reduction of $|N_0 \alpha|$, and
is expected to cross through zero for very large confinement. Rather
than a reduction, we observe an increase in $|N_0 \alpha|$ in GaMnAs
QWs. This observation is consistent with the predicted negative
contribution of the kinetic exchange, since we measure $N_0 \alpha <
0$ in our samples.

$\theta_P$ is observed to be negative for all the samples studied here
indicating that either long range Mn-Mn coupling is antiferromagnetic
or the Mn$^{2+}$ spin temperature is larger than the lattice
temperature.  Preliminary studies on modulation p-doped structures
indicate that the negative $\theta_P$ is due to a combination of
Mn$^{2+}$ spin heating by photoexcitation, as previously discussed,
and the lack of strong hole-mediated ferromagnetic Mn-Mn interaction,
which occurs for much larger $x$.  In our samples the mean Mn-Mn
distance may be too long for hole-spin coherence to be maintained,
thus precluding long-range ferromagnetic coupling.

Note that the magnetization of the Mn acceptors depends on their
electronic structure. Mn acceptors in GaAs can exist in either the
ionized $3 d^5$ (Mn$^{2+}$) configuration $A^{-}$ such that $g=2$ and
$S=5/2$, or as a neutral complex $A^{0}$ consisting of the same $3
d^5$ core with a loosely bound S=3/2 hole antiferromagnetically
coupled to it; for this entire complex $g=2.77$ and
$J=1$.\cite{Schneider:1987} Though found in GaP,\cite{Kreisel:1996}
the fully bound $3 d^4$ state with $S=4/2$ has never been observed in
GaAs. $A^{-}$ and $A^{0}$ have different angular momentum states and
thus give rise to different bulk magnetizations as measured with a
SQUID, for instance. The spin state of the core $3 d^5$ electrons
($S=5/2$), however, remains unchanged. Since $s-d$ exchange involves
the interaction between $s$ electrons in the conduction band and $d$
electrons of the Mn-ion, the value of $N_0 \alpha$ should remain
unchanged regardless of the presence of the loosely bound
hole.\cite{Szczytko:1999} In contrast, $p-d$ exchange is strongly
modified by this hole. The presence of a loosely bound hole in the
neutral complex opens a ferromagnetic exchange path whereas the
ionized acceptor offers only antiferromagnetic channels. In the
literature, such a dependence on the nature of the Mn acceptor core is
offered as an explanation for the apparent sign flip of the $p-d$ term
as the Mn concentration was increased from the very dilute limit
(paramagnetic) to the high doping regime (ferromagnetic). It may also
explain the widely varying values of $N_0 \beta$ measured in our PL
experiments (section VII). We emphasize that our exchange splitting
model takes $g_{Mn} = 2$ and thus neglects any effect of the loosely
bound hole on the core state g-factor since the exchange interaction
between the hole and the core is expected to be
small.\cite{Linnarsson:1997} Measurements in III-V DMS support
this assumption by consistently showing $g_{Mn}$ = 2.0. Therefore, the
relative concentration of $A^{-}$ and $A^{0}$ centers in our samples
should have a negligible effect on both $N_0 \alpha$ and $g_{Mn}$,
allowing us to ignore this detail in our extraction of $N_0 \alpha$
from the data.

\section {Photoluminescence}

\begin{figure}\includegraphics{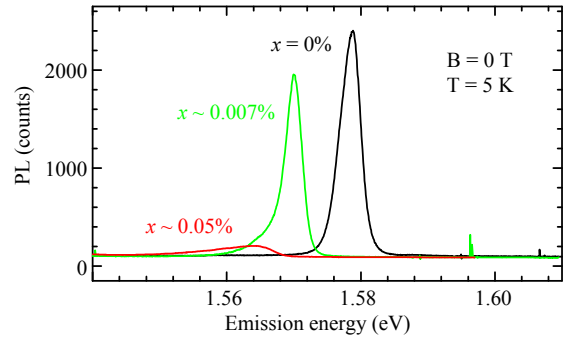}\caption{\label{fig15}
    (Color) PL plotted as a function of emission energy for a set of
    7.5-nm QWs with varying Mn-doping; the samples are excited with 1
    W/cm$^2$ at 1.722 eV. Samples with $x > 0.13$\% showed no PL.
  }\end{figure}

\begin{figure}\includegraphics[width=0.9\columnwidth]{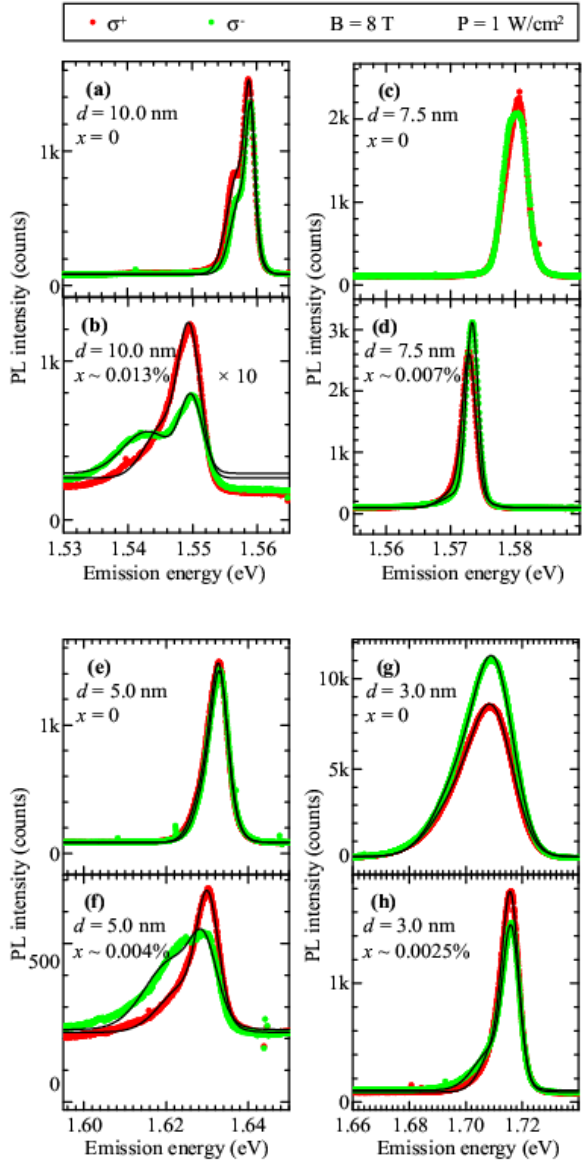}\caption{\label{fig16}
    (Color) Polarization-resolved PL for QWs of varying $d$ and $x$ at
    $ T = 5$ K.  2-Gaussian fits to the data are shown as black lines
    and the higher energy Gausssian is attributed to the heavy hole
    exciton in the QW. The excitation enregy is set to 1.722 eV for
    (a)-(f) and 2.149 eV for (g) and (h). }\end{figure}

Since hole spin lifetimes are very short in GaAs QWs ($< 10$ ps), we
rely on measurements of PL to shed light on the $p$-like valence band
and its magnetic coupling to Mn-bound $d$ electrons, $N_0 \beta$. In
addition, because recombination happens near impurities, PL can reveal
important information on defects and magnetic doping.
Polarization-resolved PL is measured as a function of $B$ in the
Faraday geometry with PL collected normal to the sample surface. The
excitation laser is linearly polarized and focused to a spot 100
$\mu$m in diameter with an energy set above the QW absorption energy.
While PL is seen to quench with increasing Mn doping, as seen in Fig.
\ref{fig15}, QWs with $x = 0$ or with small $x$ emit PL whose energy
dependence is well fit by two Gaussians (Fig. \ref{fig16}). The
emission energy of the narrower, higher-energy Gaussian peak tracks
the $B$ dependence expected for the Zeeman splitting in QWs,
indicating that this peak is due to heavy hole exciton recombination.
On the other hand, the wider, lower-energy Gaussian is likely due to
donor-bound exciton emission from shallow donors in the QWs. These
shallow donors are likely Mn$_i$, since the emission linewidth
increases as the calculated Mn$_i$ concentration increases.  Though
the lower energy Gaussian is the result of Mn doping, it is also
present in some non-magnetic samples grown with a cold Mn cell (Fig.
\ref{fig16}), perhaps due to an impurity level of Mn$_i$ ($\leq
10^{15}$ cm$^{-3}$).

In addition to quenching the PL, increased Mn-doping broadens the
low-energy emission peak. Fig. \ref{fig15} shows the zero field PL
emission at $T = 5$ K for 7.5-nm wide QWs of varying Mn-doping. The
effect of increasing Mn-doping is qualitatively identical for all QWs
of varying width: $d=$ 3.0, 5.0, 7.5, and 10.0 nm. As doping
increases, the PL broadens in energy, red shifts, and decreases in
intensity, eventually quenching. The decreasing intensity of the PL
with increasing Mn-doping parallels the degradation in KR signal with
Mn-doping. The degradation of these two optical signals, each with
distinct physical origins, i.e.  emission and absorption, reflects the
increasing density of crystalline defects with Mn-doping.

\subsection {Zeeman splitting}

\begin{figure}\includegraphics{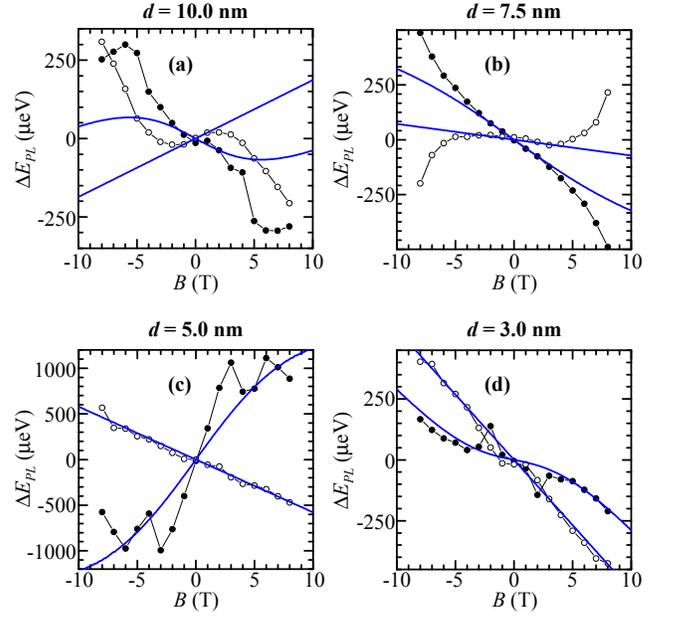}\caption{\label{fig17}
    (Color) Polarized emission splitting ($\Delta E_{PL}$) as a
    function of $B$ at $T = 5$ K for QWs of varying $d$ and $x$.  Data
    from QWs with $x = 0$ are shown as open circles, while data from
    QWs with $x \sim 0.013$\% in (a), $x \sim 0.007$\% in (b), $x \sim
    0.004$\% in (c), and $x \sim 0.0025$\% in (d) are shown as filled
    cirlces.  Fits to $\Delta E_{PL}$ for $|B| < 2$ T appear as blue
    lines.  }\end{figure}

The splitting in the polarized emission energy of the higher energy
Gaussian, $\Delta E_{PL} = E_{\sigma^{+}} - E_{\sigma^{-}}$, is
measured in all the non-magnetic samples. For small fields ($B < 2$
T), $\Delta E_{PL}$ depends linearly on field with the slope giving
the out-of-plane heavy hole exciton g-factor ($g_{ex}$). The extracted
values of $g_{ex}$ agree within the experimental error with previously
published values.\cite{Snelling:1992} At higher fields, $\Delta
E_{PL}$ deviates from linearity, particularly in the wider QWs as
shown in Fig. \ref{fig17}(a) and (b) where it reverses sign in both
the 10-nm and the 7.5-nm QWs for $x = 0$ at $|B| \sim 5$ T.

In Mn-doped samples, $\Delta E_{PL}$ results from both the Zeeman
splitting ($\Delta E_{gex}$) and the $sp-d$ exchange splitting
($\Delta E_{sp-d}$):
\begin{equation}
\label{eq6}
\Delta E_{PL} = \Delta E_{gex} + \Delta E_{sp-d} = - g_{ex} \mu_B B + x N_0 ( \alpha - \beta ) \langle S_z \rangle.
\end{equation}
Using the measurements of $g_{ex}$ from the $x=0$ samples and the
previously extracted values of $\langle S_z \rangle$ and $N_0\alpha$
at $T = 5$ K (Fig. \ref{fig11}), we fit $\Delta E_{PL}$ to Eq.
(\ref{eq6}). In the 10-nm QW for low fields we estimate $N_0
\beta=-0.85\pm0.38$ eV using the fits shown in Fig. \ref{fig17}(a) as
blue lines. As Fig.  \ref{fig17}(a) makes clear, this model breaks
down at high fields where non-linearities dominate $\Delta E_{PL}$.

Similar non-linear behavior in $\Delta E_{PL}$ at high fields in the
7.5-nm QWs, as shown in Fig. \ref{fig17}(b), contributes to the large
uncertainty in our estimates of $N_0 \beta$. Further complicating the
determination of $N_0 \beta$ are the widely differing values extracted
for samples of different widths. Using fits shown in Fig. \ref{fig17},
we find $N_0 \beta=-2.9\pm1.5$, $+24.5\pm1.8$, and $+4.3\pm0.4$ eV for
QWs with $d=$ 7.5, 5.0, and 3.0 nm, respectively.  Such dissagreement
between samples indicates the incompleteness of our model for the
valence band; the mixing of valence band states may be contributing to
the problematic extraction of the $p-d$ exchange coupling especially
for small $d$.\cite{Snelling:1992} Clearly, more work is necessary for
the determination of $N_0 \beta$ in GaMnAs QWs and its dependence on
$d$. Previous measurements in bulk GaMnAs provide little guidance with
one report suggesting positive $p-d$ exchange for low $x$
(paramagnetic) \cite{Szczytko:1996}, and others finding negative $p-d$
exchange for much larger $x$ (ferromagnetic)
\cite{Matsukura:1998,Okabayashia:1998,Zudov:2002}.

\subsection {Photoluminescence polarization}

\begin{figure}\includegraphics{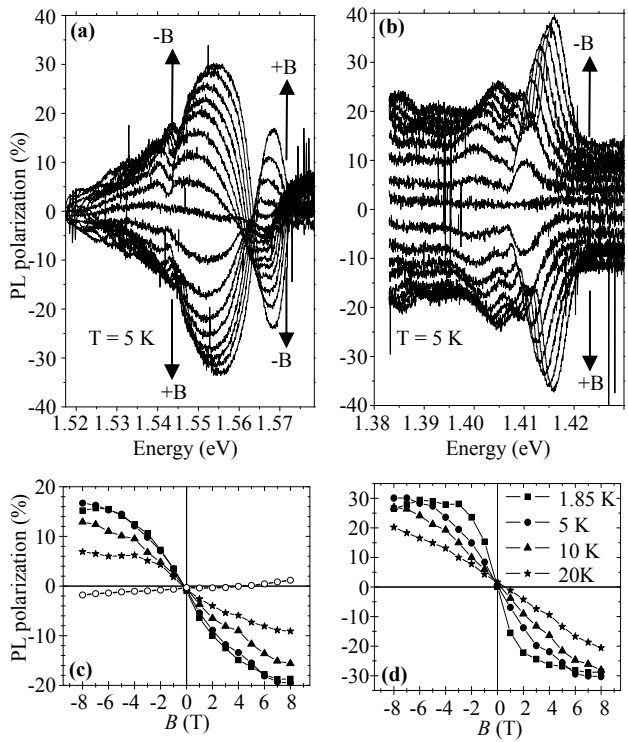}\caption{\label{fig18}
    PL polarization spectra for a 7.5-nm GaMnAs QW with an 1.72 eV
    excitation at 1.0 mW (a) near the QW emission peaks ($\sim $1.6
    eV) and (b) at the bulk Mn-acceptor line for GaMnAs ($\sim $1.4
    eV), which results from unintentional Mn-doping in the 300-nm
    thick GaAs buffer layer. The spectra are plotted as a function of
    emission energy for different magnetic fields from -8 to +8 T at 1
    T intervals. (c) The PL polarization integrated over a single
    emission peak as a function of B for the high energy ($\sim $1.57
    eV at T = 5 K, open circles) and low energy ($\sim $1.55 eV at
    several temperatures, filled symbols) peaks, and (d) for the bulk
    Mn-acceptor line peak plotted for several temperatures.
  }\end{figure}

We compare the PL polarization spectra of the 7.5-nm QW with $x \sim
0.007$\% with the well-known Mn-acceptor emission line in bulk GaAs at
1.4 eV (Fig. \ref{fig18}(a) and (b)).\cite{Yu:1979} PL polarization is
defined here as $(I_{\sigma +}- I_{\sigma -})/( I_{\sigma +}+
I_{\sigma -})$. The bulk Mn-acceptor line, shown in Fig.
\ref{fig18}(b), is measured in the same sample, resulting from the
unintentional doping of Mn in the 300-nm GaAs buffer layer grown below
the QW structure; the SIMS profiles in Fig. \ref{fig1} show that the
Mn concentration in this layer is less than $1 \times 10^{16}$
cm$^{-3}$. The polarization of this peak demonstrates a paramagnetic
(Brillouin function) field dependence, shown in Fig. \ref{fig18}(d),
following the magnetization of the Mn$_{Ga}$ acceptors in the bulk
GaAs. The low-energy peak in the QW PL polarization coincides with the
low-energy PL peak which we assigned to emission from Mn$_i$ donor
bound excitons. Its polarization is plotted in Fig. \ref{fig18}(c)
(solid points) and demonstrates similar paramagnetic (Brillouin
function) behavior with field and temperature to that of the bulk
Mn-acceptor line.

The polarization of the bulk Mn-acceptor line is proportional to the
spin-polarization of local Mn$^{2+}$ moments since the PL from this
line results from conduction band electrons recombining with holes
trapped on Mn$_{Ga}$ acceptors. The spin of these holes is coupled to
the local Mn$^{2+}$ spin.\cite{Petrou:1985} We postulate a similar
mechanism for the low-energy polarization feature in the QW PL in
which holes in the valence band of the QW recombine with electrons
bound to Mn$_i$ donors. The clear Brillouin-like field dependence
indicates that the recombinant polarization originates around isolated
paramagnetic Mn impurites in the lattice, i.e. either Mn$_{Ga}$
acceptors or Mn$_i$ donors. The Brillouin-like behavior is
inconsistent with coupled Mn centers such as
interstitial-substitutional pairs, which couple antiferromagnetically
and which are unlikely to be present in samples with such low Mn
content.\cite{Yu:2002} The $\sim 20$ meV redshift of the polarization
peak from the main QW peak does not match the 110 meV binding energy
of the substitutionial acceptor. While there is an excited state of
the Mn$_{Ga}$ acceptor with a binding energy of 26
meV,\cite{Lakshmi:2004} recombination from this excited state is
unlikely when each acceptor is filled with, at most, one hole and
given that PL usually originates from the lowest available energy
levels. Rather, we assume that the recombination originates from the
Mn$_i$ donor and the valance band in the QW. Comparison to the
experimentally measured Mn$_i$ donor binding energy is not possible
since none are reported. The postulated coupling of the electron spin
to Mn$_i$ spin results in polarized emission which follows the
magnetization of Mn$_i$ within the QW. These measurements open the
possibility of indirectly measuring the magnetization of the Mn
impurities in the QWs using polarization-resolved magneto-PL.

The high-energy feature in the QW PL polarization involves
recombination of electrons and holes bound to the QW, but delocalized
relative to the Mn states. Due to the exchange interactions in both
the valence and conduction band, the spin splitting, and thus the
resulting polarization, should have a Brillouin function field
dependence Eq. (\ref{eq6}). For the small values of $x$ studied here,
however, these effects are not resolvable and the polarization shows a
weaker field dependence Fig. \ref{fig18}(c) (open points) with an
opposite sign compared to the polarization of the low-energy peak
(solid points).

\section {Conclusions}

In summary, we demonstrate the growth of III-V GaMnAs based
heterostructures in which coherent electron spin dynamics and PL can
be observed. By lowering the substrate temperatures during growth,
sharper Mn profiles and higher Mn concentrations are attained, however
optical signals eventually quench ($T_{sub} < 400^\circ$C) likely due
to incorporation of defects by Al gettering, such as oxygen. Optical
signals are also seen to quench for $x > 0.13$\%. Hole doping due to
substitutional Mn incorporation is sensitive to the quantum well width
indicative of compensation by defects in the AlGaAs barriers. The
activation energy of holes in the QWs is lower than for an isolated
substitutional Mn in GaAs providing evidence for impurity band
formation and broadening due to large Mn doping. The crystallographic
incorporation behavior is estimated from the SIMS and Hall data and we
find that for all samples studied at least 70-90\% of Mn is located on
substitutional Ga sites. The exchange induced spin splitting in the
conduction band in the GaMnAs QWs matches the traditional paramagnetic
DMS picture and allows for the determination of the $s-d$ exchange
parameter via time-resolved electron spin spectroscopy. Suprisingly
the measured $s-d$ exchange coupling is antiferromagnetic in GaMnAs
QWs, a result not predicted by current DMS theories. No evidence of
long-range Mn spin coupling is observed, but negative effective Curie
temperatures indicate spin heating of the Mn sublattice by
photoexcitation. Electron spin lifetimes in the QWs increase for the
lowest Mn dopings compared with undoped samples indicating the
dominance of the Dyakanov-Perel mechanism over spin-flip scattering in
this regime. The ability to magnetically dope III-V and maintain
sensitive optical properties opens the door for more complex
structures to be used in the study of both free carrier and magnetic
ion spin at fast time scales, a technology which was previously
limited to II-VI DMS.

\begin{acknowledgments}
  The authors thank T. Mates (MRL, UC Santa Barbara) and Charles Evans
  and Associates for SIMS measurements, J. H. English and A. W.
  Jackson for MBE technical assistance, and J. Miller (Smelrose
  Institute) for entertaining discussions. This work was financially
  supported by DARPA, ONR, and made use of MRL Central Facilities
  supported by the MRSEC Program of the National Science Foundation
  under award No. DMR00-80034. One of us (N. P. S.) acknowledges the
  support of the Fannie and John Hertz Foundation.
\end{acknowledgments}


\begin{thebibliography}{54}
\bibitem{Ohno:1996} H. Ohno, A. Shen, F. Matsukura, A. Oiwa, A. Endo,
  S. Katsumoto, and Y. Iye, Appl. Phys. Lett. \textbf{69}, 363 (1996).
\bibitem{Dietl:2001} T. Dietl, H. Ohno, and F. Matsukura, Phys. Rev. B
  \textbf{63}, 195205 (2001).
\bibitem{Ohno:2004} H. Ohno, J. Magn. Magn. Mater. \textbf{272}, 1--6 (2004).
\bibitem{Macdonald:2005} A. H. Macdonald, P. Schiffer, and N. Samarth, Nature Materials \textbf{4}, 195-202 (2005).
\bibitem{Dietl:1994} T. Dietl, (Diluted) Magnetic Semiconductors, in
  Handbook of Semiconductors, (ed. S. Mahajan) Vol.3B (North-Holland,
  Amsterdam, 1994), p. 1251.
\bibitem{Snelling:1991} M. J. Snelling, G. P. Flinn, A. S. Plaut, R.
  T. Harley, A. C. Tropper, R. Eccleston, and C. C. Phillips, Phys.
  Rev. B \textbf{44}, 11345 (1991).
\bibitem{Snelling:1992} M. J. Snelling, E. Blackwood, C. J. McDonagh,
  and R. T. Harley, Phys. Rev. B \textbf{45}, 3922 (1992).
\bibitem{Traynor:1995} N. J. Traynor, R. T. Harley, and R. J.
  Warburton, Phys Rev. B \textbf{51}, 7361 (1995).
\bibitem{Kato3:2004} Y. K. Kato, R. C. Myers, A. C. Gossard, and D. D.
  Awschalom, Science \textbf{306}, 1910 (2004).
\bibitem{Salis:2001} G. Salis, Y. Kato, K. Ensslin, D. C. Driscoll, A.
  C. Gossard and D. D. Awschalom, Nature \textbf{414}, 619 (2001).
\bibitem{Kato:2003} Y. Kato, R. C. Myers, D. C. Driscoll, A. C.
  Gossard and D. D. Awschalom, Science \textbf{299}, 1201 (2003).
\bibitem{Poggio:2004} M. Poggio, G. M. Steeves, R. C. Myers, N. P.
  Stern, A. C. Gossard and D. D. Awschalom, Phys. Rev. B \textbf{70},
  121305(R) (2004).
\bibitem{Kato1:2004} Y. Kato, R. C. Myers, A. C. Gossard, and D. D.
  Awschalom, Nature \textbf{427}, 50 (2004).
\bibitem{Kato2:2004} Y. K. Kato, R. C. Myers, A. C. Gossard, and D. D.
  Awschalom, Phys. Rev. Lett. \textbf{93} 176601 (2004).
\bibitem{Crooker:1996} S. A. Crooker, D. D. Awschalom, J. J. Baumberg,
  F. Flack, and N. Samarth Phys. Rev. B \textbf{56}, 7574 (1997); S.
  A. Crooker, J. J. Baumberg, F. Flack, N. Samarth, D. D. Awschalom,
  Phys. Rev. Lett. \textbf{77}, 2814 (1996).
\bibitem{Myers1:2004} R. C. Myers, K. C. Ku, X. Li, N. Samarth, and D.
  D. Awschalom,  Phys. Rev. B \textbf{72}, 041302(R) (2005).
\bibitem{Johnston:2003} E. Johnston-Halperin, J. A. Schuller, C. S.
  Gallinat, T. C. Kreutz, R. C. Myers, R. K. Kawakami, H. Knotz, A. C.
  Gossard, and D. D. Awschalom, Phys. Rev. B \textbf{68} 165328
  (2003).
\bibitem{Myers2:2004} R. C. Myers, A. C. Gossard, D. D. Awschalom,
  Phys. Rev. B \textbf{69}, 161305(R) (2004).
\bibitem{Unpublished:1} Unpublished result.
\bibitem{Myers:2005} R. C. Myers, M. Poggio, N. P. Stern, A. C. Gossard, and D. D. Awschalom,  Phys. Rev. Lett. \textbf{95}, 017204 (2005).
\bibitem{Wagenhuber:2004} K. Wagenhuber, H-P Tranitz, M. Reinwald, and
  W. Wegscheider, Appl. Phys.Lett. \textbf{85}, 1190 (2004).
\bibitem{Nazmul:2003} A. M. Nazmul, S. Sugahara, and M. Tanaka, J.  Crys. Grow. \textbf{251}, 303 (2003). 
\bibitem{Akimoto:1986} K. Akimoto, M. Kamada, K. Taira, M. Arai, and N. Watanabe, J. Appl. Phys. \textbf{59}, 2833 (1986).
\bibitem{Achtnich:1987} T. Achtnich, G. Burri, M. A. Py, and M. Ilegems, Appl. Phys. Lett. \textbf{50}, 1730 (1987).
\bibitem{Snider:1} 1D Poisson-Schroedinger solver written by G. Snider, (http://www.nd.edu/~gsnider/).
\bibitem{Toyoshima:1993} H. Toyoshima, T. Niwa, J. Yamazaki, and A. Okamoto, Appl. Phys. Lett. \textbf{63}, 821 (1993).
\bibitem{Nagle:1993} J. Nagle, J. P. Landesman, M. Larive, C. Mottet, and P. Bois, J. Cryst. Growth \textbf{127}, 550 (1993).
\bibitem{Macguire:1987} J. Maguire, R. Murray, R. C. Newman, R. B. Beall, and J. J. Harris, Appl. Phys. Lett. \textbf{50}, 516 (1987).
\bibitem{Grandidier:1998} B. Grandidier, D. Stiévenard, J. P. Nys, and X. Wallart, Appl. Phys. Lett. \textbf{72}, 2454 (1998).
\bibitem{Missous:1994} M. Missous and S. O'Hagan, J. Appl. Phys. \textbf{75},
  3396 (1994).
\bibitem{Yu:1979} P. W. Yu and Y. S. Park, J. Appl. Phys. \textbf{50},
  1097 (1979).
\bibitem{Blakemore:1973} J. S. Blakemore, Winfield J. Brown, Jr., Merrill L. Stass, and Dustin A. Woodbury, J. Appl. Phys. \textbf{44}, 3352 (1973).
\bibitem{Woodbury:1973} D. A. Woodbury and J. S. Blakemore, Phys. Rev. B \textbf{8}, 3803 (1973).
\bibitem{Erwin:2002} S.C. Erwin and A.G. Petukhov, Phys. Rev. Lett.
  \textbf{89}, 227201 (2002).
\bibitem{Zhao:2005} L. X. Zhao, C. R. Staddon, K. Y. Wang, K. W. Edmonds, R. P. Campion, B. L. Gallagher, and C. T. Foxon, Appl. Phys. Lett. \textbf{86}, 071902 (2005).
\bibitem{Yu:2002} K. M. Yu, W. Walukiewicz, T. Wojtowicz, I. Kuryliszyn, X. Liu, Y. Sasaki, J. K. Furdyna, Phys. Rev, B \textbf{65}, 201303(R) (2002).
\bibitem{Raebiger:2004} H. Raebiger, A. Ayuela, and R. M. Nieminen, J. Phys.: Condens. Matter. \textbf{16}, L457 (2004).
\bibitem{Fabian:1999} J. Fabian and S. das Sarma, J. Vac. Sci. Technol. B \textbf{17}, 1708 (1999).
\bibitem{Optical:1984} \textit{Optical Orientation, Modern Problems in Condensed Matter Science}, edited by F. Meier and B. P. Zachachrenya (North-Holland, Amsterdam, 1984), Vol. 8.
\bibitem{Semenov:2003} Y. G. Semenov, Phys. Rev. B \textbf{67}, 115319 (2003).
\bibitem{Keller:2001} D. Keller, D. R. Yakovlev, B. Konig, W. Ossau,
  Th. Gruber, A. Waag, and L. W. Molenkamp, Phys. Rev. B \textbf{65},
  035313 (2001).
\bibitem{Weisbuch:1977} C. Weisbuch and C. Hermann, Phys. Rev. B
\textbf{15}, 816 (1977).
\bibitem{Larson:1988} B. E. Larson, K. C. Hass, H. Ehrenreich, and A.
  E. Carlsson, Phys. Rev. B \textbf{37}, 4137 (1988).
\bibitem{Merkulov:1999} I. A. Merkulov, D. R. Yakovlev, A. Keller, W.
  Ossau, J. Heurts, A. Waag, G. Landwehr, G. Karczewski, T. Wojtowicz,
  and J. Kossut, Phys. Rev. Lett. \textbf{83}, 1431 (1999).
\bibitem{Schneider:1987} J. Schneider, U. Kaufmann, W. Wilkening, M. Baeumler, Phys. Rev. Lett. \textbf{59}, 240 (1987).
\bibitem{Kreisel:1996} J. Kreisel, W. Ulrici, Phys. Rev. B \textbf{54} 10508 (1996).
\bibitem{Szczytko:1999} J. Szczytko, W. Mac, A. Twardowski, F. Matsukura, H. Ohno, Phys. Rev. B \textbf{59}, 12935 (1999).
\bibitem{Linnarsson:1997} M. Linnarsson, E. Janzen, B. Monemar, M. Kleverman, A. Thilderkvist, Phys. Rev. B \textbf{55}, 6938 (1997).
\bibitem{Szczytko:1996} J. Szczytko, W. Mac, A Stachow, A Twardowski, P. Becla,
and J. Tworzydlo, Solid State Commun. \textbf{99}, 927 (1996).
\bibitem{Matsukura:1998} F. Matsukura, H. Ohno, A. Shen, and Y. Sugawara, Phys.
Rev. B \textbf{57}, 2037(R) (1998).
\bibitem{Okabayashia:1998} J. Okabayashia, A. Kimura, O. Rader, T. Mizokawa, 
A. Fujimori, T. Hayashi, and M. Tanaka, Phys. Rev. B \textbf{58}, R4211 (1998).
\bibitem{Zudov:2002} M. A. Zudov, J. Kono, Y. H. Matsuda, T. Ikaida, N. Miura, 
H. Munekata, G. D. Sanders, Y. Sun, and C. J. Stanton, Phys. Rev. B 
\textbf{66}, 161307(R) (2002).
\bibitem{Petrou:1985} A. Petrou, M. C. Smith, C. H. Perry, J. M.
  Worlock, J. Warnock, and R. L. Aggarwal, Sol. State Commun.
  \textbf{55}, 865 (1985).
\bibitem{Lakshmi:2004} B. Lakshmi, G. Favrot, D. Heiman, Proc. SPIE \textbf{5359}, 290 (2004).
\end{thebibliography}
\end{document}